%% file: _main.tex
\documentclass[acmsmall]{acmart}

\usepackage{multirow}
\usepackage{xcolor}
\usepackage{longtable}
\usepackage{rotating}
\usepackage{placeins}
\usepackage{float}

\AtBeginDocument{%
  }

\setcopyright{cc}
\setcctype{by}
\acmJournal{PACMHCI}
\acmYear{2027} \acmVolume{11} \acmNumber{1} \acmArticle{GROUP23}
\acmMonth{1} \acmDOI{10.1145/3859409}

\begin{document}

\title[Harassment on TikTok]{Harassment on TikTok: A Mixed--Methods Analysis of Video Content and Psycholinguistic Comment Stances}



\author{Afsaneh Razi}
\affiliation{%
  \institution{Drexel University}
  \city{Philadelphia}
  \state{PA}
  \country{USA}
}
\email{afsaneh.razi@drexel.edu}
\orcid{0000-0001-5829-8004}

\author{Tyler Chang}
\affiliation{%
  \institution{Drexel University}
  \city{Philadelphia}
  \state{PA}
  \country{USA}
}
\email{tlc373@drexel.edu}
\orcid{0009-0002-6498-2827}

\author{Michal Monselise}
\affiliation{%
  \institution{Drexel University}
  \city{Philadelphia}
  \state{PA}
  \country{USA}
}
\email{mm4837@drexel.edu}
\orcid{0000-0002-6222-4822}

\author{Joseph J. Trybala}
\affiliation{%
    \institution{University of Colorado Boulder}
    \city{Boulder}
    \state{CO}
    \country{USA}
}
\email{Joseph.Trybala@colorado.edu}
\orcid{0009-0007-2379-8683}

\author{Asta Zelenkauskaite}
\affiliation{%
    \institution{Drexel University}
    \city{Philadelphia}
    \state{PA}
    \country{USA}
}
\email{az358@drexel.edu}
\orcid{0000-0001-5762-4605}

\begin{abstract}
    \input{sections/0-Abstract}

\end{abstract}



\maketitle
\input{sections/1-Introduction}

\input{sections/2a-New_Related_Works}
\input{sections/3-Methods_new}
\input{sections/4a-new_results}

\input{sections/5-Discussion}
\input{sections/6-Conclusion}

\section*{AI Usage Disclosure} GPT-5.2 was used to resolve minor grammatical errors and improve table formatting. Google Scholar Labs was used to identify some of the supporting literature. 

\section*{Ethics and Privacy Statement} We do not expect any significant societal risks to result from the publication of this work. Special care was taken to minimize the chances of any of the examined users' being identified from their inclusion in our analysis. We anticipate that our findings, if considered by TikTok moderators, would require a greater investment into developing automated processes for identifying potentially harassment-including content. We also expect that such adoptions would improve but not fully resolve the issue of malicious harassment perpetrated through the platform.

\section*{Acknowledgments} This paper is supported in part by the U.S. National Science Foundation under grant \#2542768. Any opinions, findings, and conclusions or recommendations expressed in this material are those of the authors and do not necessarily reflect the views of the research sponsor.

\bibliographystyle{ACM-Reference-Format}
\bibliography{references.bib,ref-zotero-ar}
\received{31 March 2026}
\received[revised]{17 July 2026}
\received[accepted]{17 September 2026}

\appendix
\input{sections/7a-new_appendix}

\end{document}

%% file: sections/0-Abstract.tex
Despite extensive research on harassment, how it manifests on short--form video platforms such as TikTok---and how it shapes user interactions---remains underexplored, though these dynamics are critical for designing effective safety strategies for online communities. We present a mixed-methods study of 190 TikTok videos tagged with \#harassment and their 66,337 comments. Our qualitative analysis of videos identifies harassment types (sexual, verbal, stalking, law enforcement, bullying, catcalling, and gaslighting) and their characteristics (e.g., location, target, video style, and theme), along with engagement patterns. Although sexual harassment was the most prevalent topic, it did not have the highest engagement; instead, stalking--related videos received the most views. Pairwise comparisons and regression--based analyses of comments’ psycholinguistic features and stance (support, opposition, neutrality) revealed differences in emotional tone, social orientation, and cognitive categories across harassment types. Comments on sexual harassment frequently express solidarity with victims, whereas discussions of law enforcement harassment emphasize institutional critique. We provide implications of our findings, demonstrating that harassment on TikTok is multifaceted, shaped by content, context, and community norms.

%% file: sections/1-Introduction.tex
\section{Introduction}

TikTok, with over a billion users---most under 30 \cite{laughing}---is a powerful platform for digital expression and public discourse. Prior research shows that social media platforms 
shape public conversations on a variety of issues \cite{basch2022deconstructing}. However, this influence also creates risks, including the amplification of online harassment, defined as behaviors enacted through electronic means with the intent to harm a victim \cite{slaughter_new_2022}. In particular, TikTok's platform affordances---including its algorithmically driven "For You" page, comment--centric engagement patterns, and the visibility dynamics of short--form videos that foreground creators’ identities and reactions~\cite{devito2022transfeminine,are2023autoethnography}---can facilitate the spread and visibility of harassment. This is particularly concerning given that 41-47\% of internet users aged 15-18 report experiencing aggression online, including forms of harassment \cite{slaughter_new_2022,vogels_state_2021}.

CSCW and GROUP scholars have long examined online harassment as a sociotechnical phenomenon, emphasizing how platform affordances, community norms, and interactional dynamics shape both harmful behavior and supportive responses. Prior work has studied harassment across several contexts in online communities~\cite{take_stoking_2024}, gaming environments, virtual reality (VR)~\cite{freeman_i_2025,zheng_understanding_2023}, and social media~\cite{razi2023sliding}, and various populations including female gamers \cite{mclean_female_2019}, transfeminine content creators~\cite{devito2022transfeminine}, and adolescents \cite{beran_evidence_2012, razi2023sliding}. They have highlighted issues such as bystander intervention, networked harassment, and the role of moderation and governance in shaping discourse~\cite{jhaver2019Online}.
In parallel, researchers have also investigated automatic detection methods~\cite{kim2021human,Razi2021KIm} and online moderation techniques~\cite{jhaver2019Online,zeng2022content,lyu2024got} to identify and mitigate online harassment and cyberbullying. 
Studies on harassment types and larger comment analysis on social media platforms have focused primarily on textual modality~\cite{razi2023sliding,kim_when_2022}, and some have focused on images~\cite{ali_getting_2023}; however, understanding harassment on short--form video platforms remains limited. 

More recent studies have begun to examine harassment on short--form video platforms, focusing on content creator experiences and their relation to experiencing harassment~\cite{devito2022transfeminine}, algorithmic visibility~\cite{de_whoever_2025}, and attention and fame on TikTok~\cite{schafer_i_2025}. However, these efforts have largely centered on content creators or platform design based on self-reported research methods, with comparatively limited attention to systematically characterizing harassment types and their downstream interactional effects in these environments.
Therefore, to better understand how harassment discourse manifests in algorithmically curated short--form video environments, we ask the following research questions:
\begin{description}
\item \textbf{RQ1:} \textit{What types of harassment appear in TikTok videos surfaced through the \#harassment hashtag, and how do their contextual characteristics vary?}

\item \textbf{RQ2:} \textit{How do comments responding to different harassment types differ psycho--linguistically?}

\item \textbf{RQ3:} \textit{a) How does the type of harassment depicted in a video influence whether commenters express neutral, supportive, or oppositional stances? b) How do psycholinguistic features differentiate supportive from oppositional engagement in responses to harassment content?}
\end{description}

Using qualitative content analysis, Linguistic Inquiry and Word Count (LIWC), and stance detection, we analyzed 190 videos and their 66,337 associated comments. Building on a statistically robust framework, including pairwise comparisons (Dunn's test) and a Bayesian hierarchical multinomial logistic regression model, we identify the linguistic properties that differentiate types of harassment and characterize the stances of comments, both supportive and oppositional. In doing so, we found that harassment on TikTok is multifaceted, contrasting with prior work that treats it as a single phenomenon. Moreover, comments responding to different types of harassment exhibit distinct linguistic and emotional response patterns, suggesting that commenters adopt different rhetorical styles depending on the context of the harassment. Finally, while commenters are generally neutral regarding harassment, when comments are polarized, they are overwhelmingly supportive, though this variation depends more on the idiosyncrasies of individual videos than on harassment type.
Our work makes the following research contributions:

\begin{itemize}
    \item A contextual characterization of harassment narratives in algorithmically served TikTok videos that moved beyond self-reported research methods to understand the content.
    \item A large--scale psycholinguistic examination of audience comments to harassment allowed us to move beyond surface--level engagement metrics and capture how linguistics, emotional tone, social orientation, and cognitive processes vary across different forms of harassment.
    \item A stance--based analysis of how TikTok audiences express supportive, neutral, and oppositional engagement with harassment narratives, which builds on research on bystander behavior and collective sensemaking, offering insight into how users respond to harmful content.
\end{itemize}

In doing so, this work provides insight into how harassment is represented, interpreted, and debated within short--form video environments. Consequently, these findings contribute to a more nuanced understanding of harassment discourse on TikTok and highlight the importance of incorporating both contextual and linguistic perspectives when studying harmful online interactions.

%% file: sections/2a-New_Related_Works.tex
\section{Background and Related Works}
\label{sec:background_and_related}

\subsection{Online Harassment on Social Media}

Online harassment has been widely studied as a form of harmful behavior enacted through digital communication platforms. Prior research has examined diverse aspects of this phenomenon, including patterns of online abuse \cite{WomansPrespect2022,OnlineHarassment2023}, survivor narratives and support--seeking behaviors \cite{andalibi2016,heidi2021,razi2020lets}, and large--scale social movements such as \#MeToo \cite{MeinMetoo2021,BangladeshiMetoo2020}. These works highlight how social media platforms both expose individuals to harassment and enable victims to share experiences, mobilize support, and advocate for social change. A substantial portion of the literature has also focused on computational approaches for detecting harassment and abusive behavior online \cite{liuYing2019,razi2023sliding,Razi2021KIm,khatua2020,kim2021human,ghosh-chowdhury-etal-2019-speak}. These studies examine linguistic patterns associated with harmful content and develop automated methods to identify harassment in large--scale datasets. Research has also documented the psychological consequences of online abuse, linking repeated exposure to outcomes such as post--traumatic stress disorder (PTSD) and emotional distress \cite{McHugh2018WhenSM, phillips2018can}. In particular, Phillips has emphasized that recovery from harassment--induced trauma lacks a quick resolution, often requiring victims to "re--cognize" their own self-worth after the active harassment ends \cite{phillips2018can}. On the other hand, scholars have noted that social media platforms can amplify marginalized voices and enable collective responses to harassment through sociopolitical campaigns and community discourse \cite{Leopold2019TheHH}. Despite these advances, much of the existing literature tends to treat harassment as a vaguely defined or monolithic category. Less attention has been paid to how different types of harassment---such as sexual harassment, verbal harassment, and cyberbullying---may differ in their contextual framing, narrative structure, and audience response. This gap is particularly salient on emerging platforms like TikTok, where short--form video content and algorithmic recommendation systems shape how harassment--related content is produced, circulated, and interpreted.

\subsection{Classifying Harassment and Harmful Content}

Despite the prominence of online harassment as a research subject, it lacks a unified definition as a category. While it is most often associated with sexuality-- or gender--related topics \cite{henry2018technology, Razi2021KIm}---where it frequently encompasses both verbal and physical forms of violence---its scope is considerably broader in practice. Several researchers have explicitly incorporated negative social or emotional consequences into their definitions \cite{nova2023integrating, blackwell2017classification}, emphasizing behaviors such as shaming and insulting others. Other scholars \cite{vaillancourt2017cyberbullying} employ semantically related terms---such as bullying and cyberbullying---to broadly capture any form of deliberately inflicted, digitally mediated social or emotional harm. In contrast, some researchers emphasize distinctions between cyberbullying and online harassment across different contexts. For example, Kim et al. \cite{kim_human-centered_2021} highlight the frequent omission of key characteristics---such as repetition, malicious intent, and power imbalance---in many historical definitions of cyberbullying. This lack of conceptual consistency has motivated efforts to more systematically classify and categorize harmful online behaviors.

Several studies have attempted to classify or categorize harmful behaviors in online environments. For example, Rezvan et al. \cite{rezvan_analyzing_2020} examined online verbal harassment and proposed five categories of abusive behavior: sexual, racial, appearance--related, intellectual, and political harassment. Such taxonomies provide useful starting points for understanding the range of abusive behaviors that occur in online discourse. However, many existing classification frameworks were developed using text--based communication platforms and may not fully capture the complexity of multimedia environments. Video--based content can contain multiple themes, narrative styles, and interactional cues within a single post \cite{yusoff_thematic_2023}. Moreover, some harassment categories that are prominent in text--heavy platforms such as Twitter (X)---for example, political or race--related harassment---may manifest differently in short--form video environments like TikTok \cite{gelles-watnick_how_2024}. Other scholars have proposed broader frameworks for categorizing harmful online content beyond harassment alone. For instance, Scheuerman et al. \cite{scheuerman_framework_2021} developed a framework for classifying harmful online interactions that accounts for contextual factors and social dynamics surrounding abusive behavior. In addition, some studies have employed inductive approaches to classification, combining qualitative coding with interviews or participant feedback to identify emergent categories of online harm. Building on this work, our study adopts an inductive approach to identifying harassment categories within TikTok videos. Through manual coding and intercoder reliability checks, we identify platform--specific forms of harassment representation that reflect the narrative and thematic characteristics of short--form video content.

\subsection{Contextual and Psycholinguistic Analysis of Harassment Discourse}

Most research on detecting online harassment has focused on textual features that capture linguistic patterns within user--generated content \cite{Razi2021KIm}. While these approaches can effectively identify indicators of abusive language, they often emphasize \textit{what} is being said while overlooking \textit{how} harassment narratives unfold and how audiences interpret them. Understanding these contextual dynamics is important because individuals interpret online content differently based on their social identities, experiences, and relationships to the content being discussed \cite{kim2021human}. Prior studies have examined how contextual factors shape harassment discourse across platforms---for example, how tone and perceived threat levels differ between political and gender--based abuse on Twitter \cite{rheault2019line}, or how relational dynamics influence the interpretation of cyberbullying among youth \cite{marwick2014s}. In parallel, psycholinguistic approaches such as sentiment analysis and tools like Linguistic Inquiry and Word Count (LIWC) have been used to analyze emotional and cognitive markers in online discussions surrounding trauma, support--seeking, and harmful experiences \cite{andalibi2016}. These approaches provide insight into how language reflects users’ emotional states, coping strategies, and social responses when engaging with distressing content. Extending these approaches, our study examines whether different types of harassment are associated with distinct contextual patterns in video content and psycholinguistic differences in user comments. Identifying such variation can provide insight into how harassment narratives are framed and how audiences interpret and respond to them.

\subsection{Short-Form Video Platforms and Harassment Discourse}

Short--form video platforms introduce distinct affordances that shape how online discourse unfolds. TikTok’s algorithm--driven “For You” page curates content based on user engagement and viewing patterns, connecting individuals to communities and conversations organized around shared interests and hashtags \cite{TikToktradeoff}. These recommendation mechanisms can facilitate the rapid spread of narratives and enable the formation of \textit{ad hoc} communities centered on particular topics. Recent research has shown that TikTok plays a significant role in shaping public discussions about a range of social issues, including mental health \cite{Zhang2023TikTok, Milton2023}, political discourse \cite{polarizationTikTok, politics}, and science communication \cite{science, covidresponnse}. For example, Milton et al. introduced the concept of “permeability” to describe how TikTok’s open content boundaries allow users to exchange support and explore identities across different communities \cite{Milton2023}. These communities, which emerge both from users deliberately searching for like--minded individuals and from algorithmically mediated clustering \cite{gerbaudo2024tiktok, kang2022ai}, therefore structure not only what users are exposed to but also how they interpret content and their willingness to disclose personally sensitive experiences. Moreover, despite efforts to control the content to which they are exposed and, consequently, the representation of their digital selves on TikTok, users cannot exert this control to their desired degree \cite{simpson2022tame}. As a result, users have only a limited capacity to avoid or disengage from harassment content, except by ceasing their use of TikTok altogether.

Despite the growing influence of TikTok, research on harassment discourse within this environment remains limited. Compared to platforms such as Twitter---where harassment--related discourse and activism have been extensively studied---TikTok’s video--centric format and algorithmic content distribution create distinct dynamics that are not yet well understood. Prior work has applied topic modeling to TikTok comments related to harassment and identified themes such as law enforcement responses, self-defense strategies, and community reactions \cite{armin2024support}. However, relatively little research has examined how harassment--related content is algorithmically surfaced on the platform or how audience responses vary linguistically across different types of harassment. Our study addresses this gap by investigating how harassment is represented in videos surfaced through the \#harassment hashtag on TikTok and how audiences respond through psycholinguistic patterns in comment discourse.

%% file: sections/3-Methods_new.tex
\section{Methods}

Section~\ref{sec:data_collect} explains the data collection and refinement process, including ethical considerations. Section~\ref{sec: qualitative_analysis_of_videos} describes the qualitative coding process applied to the dataset. Section~\ref{sec: psycholinguistic_analysis_of_videos} recounts our use of LIWC to identify the psycholinguistic features that differentiated the comments of various types of harassment. Finally, Section~\ref{sec:methods_stance} explicates our application of stance detection on the comments to assess reactionary polarization.

\subsection{Data Collection and Refinement}
\label{sec:data_collect}

This study employed video and comment--level analyses to assess harassment types and psycholinguistic patterns. We first collected 884 videos tagged with \#harassment in the title or description on January 8, 2024 using an unofficial TikTok API that interfaces with TikTok's web API \footnote{\url{https://github.com/davidteather/TikTok-Api}}. The choice to use an unofficial API was strictly determined due to the inaccessibility of the official API at the time. Although the data collection was limited to a single day, the dates on which the videos were posted were not similarly restrained. While all videos were published prior to the collection date, we did not impose a recency constraint; the determination of whether a video was considered for analysis was based solely on whether the unofficial TikTok API surfaced it when searching for content tagged with \#harassment. Our use of hashtag--based searching is much akin to that of \cite{Feuston2019everyday}, where a similar method was used to collect Instagram videos relating to mental health disclosures. As TikTok's content recommendation system is proprietary and therefore opaque, we treated these videos as a snapshot of popular content fitting our criteria at the time. Relatedly, since all of the video authors were based in the United States, we assumed that the videos captured were representative of content served to U.S. users, rather than reflective of TikTok as a whole. We also necessarily considered our data to be exploratory on account of the sheer volume of content uploaded to and removed from TikTok each day. For each video, we gathered its URL, associated comments, and engagement metrics (likes, shares, saves, views). All videos were publicly accessible at the time of collection, though some have since been made unavailable. 

We removed videos 
if they lacked comments, had majority non-English comments, or had been removed prior to our review of the videos.
The language--based exclusion criterion was necessary to allow all authors to review and interpret the comments. 
This resulted in the removal of 13 videos that lacked any comments. After confirming there were no duplicates, we verified that all remaining videos explicitly included the \#harassment hashtag. While we did not assume that tag guaranteed the video's relevance to harassment, we retained all videos that met the hashtag criterion.


\subsubsection{Ethical Considerations}
All the gathered content was public information since it could be accessed without logging into the website. Although IRB considers this activity as not human subjects research, considering the ethics of using publicly available data for research is crucial. 
While it is true that the data is accessible without an account, this does not necessarily mean that content creators or commenters anticipated their contributions being used in a study \cite{fiesler2018participant}. We therefore ensured that no personally identifiable information (PII) was included in our results. Relatedly, we lightly edited any comments referenced in this paper to ensure that sensitive disclosures were omitted and to better protect the privacy of the comments' authors.

\clearpage
\begin{figure*}[p]
\centering
\rotatebox{90}{%
\begin{minipage}{0.95\textheight}
\centering
\includegraphics[
    width=\linewidth,
    height=0.95\textwidth,
    keepaspectratio,
    alt={A visualization of the data collection, cleaning, annotation, and linguistic analysis methods}
]{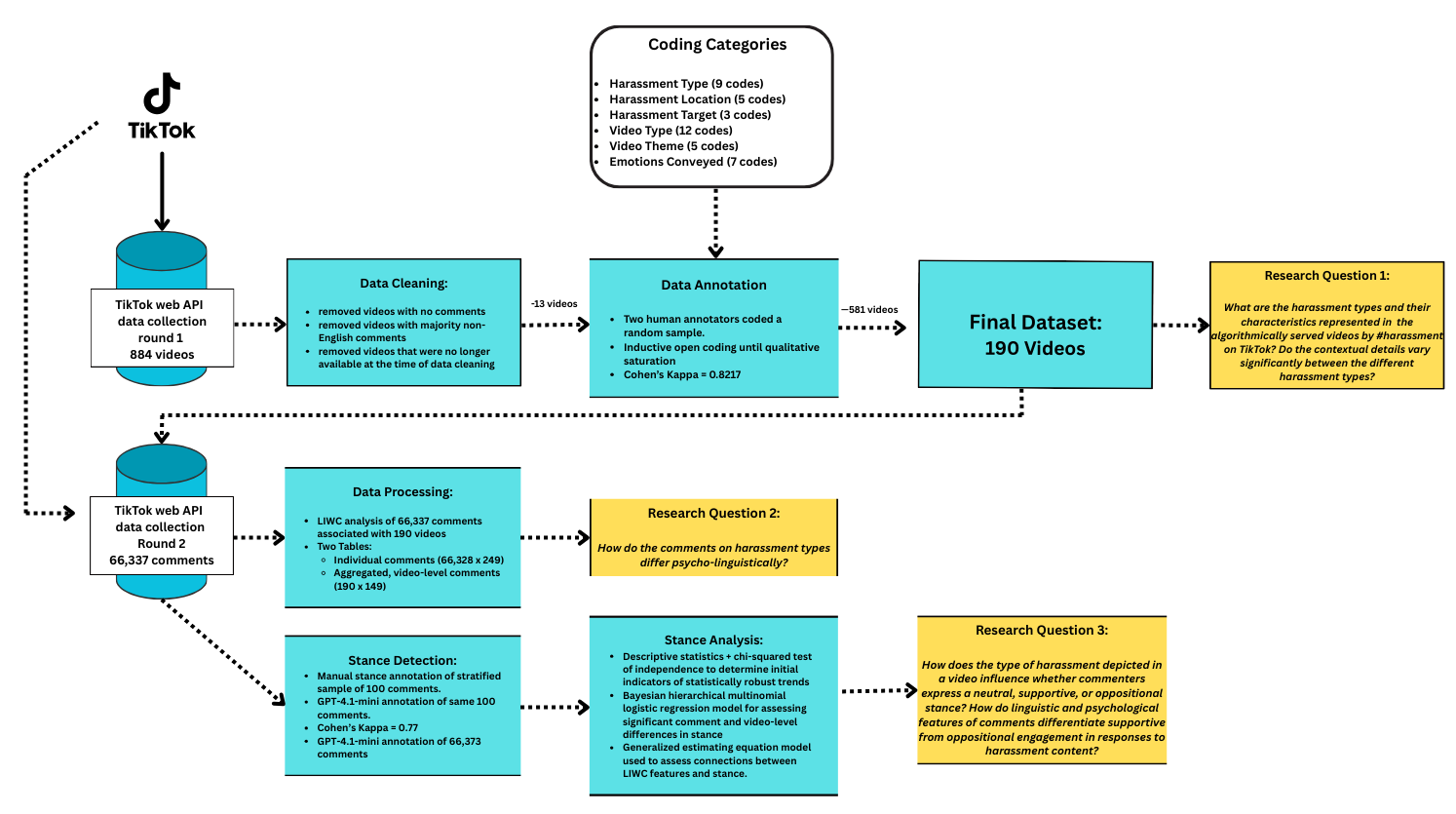}
\caption{Describes the process of collecting, cleaning, annotating, and psycholinguistically evaluating the 190 harassment-related videos and the 66,337 associated comments from TikTok.}
\label{fig:methods_figure}
\end{minipage}
}
\end{figure*}
\clearpage

\subsection{Qualitative Analysis of Videos}
\label{sec: qualitative_analysis_of_videos}
We randomly sampled videos and open coded them iteratively until no new codes emerged, yielding a final sample of 190 videos suitable for qualitative content analysis~\cite{krippendorff2018content}.
Two researchers independently conducted inductive open coding of a 10 percent subset of the videos and then collaboratively developed a unified, bottom--up codebook (Table~\ref{tab:codebook_flat}). Then, each researcher coded all 190 videos, achieving a mean Cohen's Kappa of 0.8217 across coding features, indicating strong agreement; disagreements were resolved through discussion.
After final coding, we used LIWC to analyze the 
emotional and linguistic qualities of associated comments by breaking down videos based on harassment type. We also calculated code prevalence in anticipation of our quantitative analysis. Our codes included:

\begin{itemize}
    \item \textbf{Harassment Type:} The type of harassment that appears in the video. This includes verbal
, sexual
, physical
, stalking
, bullying
, catcalling
, law enforcement
, gaslighting
    , and other.
    \item \textbf{Harassment Location:} Where the harassment took place. It includes in person
    , online
    , scripted
    , in a movie, and other as locations.
    \item \textbf{Harassment Target:} Who was being harassed. This category served to differentiate videos where the harassment was self--reported from those in which the harassment was reported on behalf of someone other than the uploader. The codes includes self
    , other
    , and celebrity.
    \item \textbf{Video Type:} The style of video. It includes point-of-view
    , documentary
    , oral speech
    , news
    , acting
    , animated infographic
, responding to stitch
    , talking over background image
    , picture slideshow
    , cartoon
    , and other.
    \item \textbf{Video Theme:} The primary theme or focus of the video.  This includes knowledge/awareness
    , personal experience
    , personal precaution
    , news
    , and other.
    \item \textbf{Emotion Conveyed:} The primary emotion of the person filming the video. This includes alarm/concern
    , news
    , humor
    , response efficacy
    , hope
    , normal life impacts
    , and other.
\end{itemize}



\begin{table}[h]
\scriptsize
\centering
\resizebox{\textwidth}{!}{%
\begin{tabular}{lll}
\toprule
\textbf{Dimension} & \textbf{Code} & \textbf{Code Description} \\
\midrule
Harassment Location & In person (n = 150, 78.95\%) & The video describes a type of harassment that happened in a face--to--face interaction.\\
 & Online/Over Message (n = 20, 10.53\%) & The video describes harassment that happened either in an online exchange or via text message.\\
 & Unrealistic Scenario (n = 15, 7.89\%) & \\
 & $\bullet$ \textit{Scripted (n = 12, 6.32\%)} & The video describes a scripted interaction in a video made for TikTok rather than a clip from a film.\\
 & $\bullet$ \textit{In a movie (n = 3, 1.58\%)} & The video shows a scripted scene that portrays harassment in a film.\\
 & Other (n = 5, 2.63\%) & A location that cannot be classified as any of the other location categories.\\
\midrule

Harassment Target* & Self (n = 97, 51.07\%) & The person in the video is describing or documenting an incident of harassment that happened to them.\\
 & Other (n = 84, 44.21\%) & The person in the video is describing or documenting an incident of harassment that happened to someone else.\\
 & Celebrity (n = 9, 4.74\%) & The video describes or displays an incident of harassment that happened to a celebrity.\\
\midrule

Harassment Type* & Sexual (n = 120, 63.16\%) & \\
 & $\bullet$ \textit{Inappropriate Sexual Behavior (n = 94, 49.47\%)} & Any unwanted, uninvited or inappropriate behavior of a sexual nature, including verbal comments, touching,\\ & & patting, hugging or other physical contact \cite{rapecrisisWhatSexual, eeocSexualHarassment}. \\
 & $\bullet$ \textit{Catcalling (n = 26, 13.68\%)} & Street harassment that involves making an inappropriate comment to a passerby \cite{stopstreetharassmentDefinitionsStop}.\\
 & Verbal (n = 44, 23.16\%) & \\
 & $\bullet$ \textit{Verbal Harassment (n = 37, 19.47\%)} & Shouting, insulting, threatening, or otherwise verbally assaulting another person \cite{madialawWhatVerbal}.\\
 & $\bullet$ \textit{Gaslighting (n = 7, 3.68\%)} & An interaction where the person is being manipulated either in person or online into questioning their sanity \\ & & or own ability to use reason \cite{Sweet2019socio}.\\
 & Bullying (n = 44, 23.16\%) & An interaction where the attacker seeks to harm or coerce the other person \cite{cdcBullying}.\\
 & Stalking/Staring (n = 25, 13.16\%) & An interaction where the person is being followed in a physical or online location \cite{Spitzberg2007state}.\\
& Law Enforcement (n = 15, 7.89\%) & An interaction where the attacker is acting in their capacity as a member of a law enforcement agency \cite{legalclarityWhatConsidered}.\\
 & Physical (n = 10, 5.26\%) & Physically harming someone in a non--sexual way, including punching, hitting, pushing, slapping, \\ & & kicking, biting, or tripping another person \cite{womenshealthPhysicalAbuse}.\\
 & Other (n = 9, 4.74\%) & An interaction that cannot be classified as one of the other harassment type categories.\\
 \midrule
 Video Type & Oral Speech (n = 55, 28.95\%) & A video that primary consists of an individual talking to the camera and describing an event that happened \\ & & in the past or their opinion on a topic.\\
 & Other (n = 40, 21.06\%) & A video type that cannot be classified as one of the other video type categories.\\
 & Point-of-View (n = 32, 16.84\%) & A video where the person filming is showing an incident or interaction from their point--of--view, typically \\ & & exclusively filming others and not themselves.\\
  & Responding to Stitch (n = 24, 12.63\%) & A video where another video is first shown and then the person who uploaded the video responds to that \\ & & video with their own opinion.\\
& Informative (n = 20, 10.53\%) & \\
 & $\bullet$ \textit{Acting (n = 12, 6.32\%)} & A video in which information is presented via individual or group acting or as role play.\\
 & $\bullet$ \textit{News (n = 8, 4.21\%)} & A video that presents information about recent or important events or newly released statistics.\\
 & Talking over Background Image (n = 17, 8.95\%) & A video where the speaker displays an image in the background and discusses the image.\\
 & Documentary (n = 3, 1.58\%) & A video that provides a factual record or report of events or people.\\
 \midrule
Video Theme* & Personal Experience (n = 101, 53.16\%) & A video where the victim describes an incident that happened to them in the past or is showing the \\ & & incident happening in real time.\\
 & Knowledge/Awareness (n = 45, 23.68\%) & A video that presents information about the types, prevalence, severity, susceptibility, or any other \\ & & aspect of harassment to inform viewers.\\
 & Personal Precaution (n = 23, 12.11\%) & A video that focuses on the precautionary measures taken by individuals to avoid or prevent harassment.\\
 & Other (n = 15, 7.89\%) & A video them that cannot be classified as one of the other video theme categories.\\
 & News (n = 6, 3.16\%) & A video that describes incidents of harassment in a news article.\\
\midrule
Emotions Conveyed* & Alarm/Concern (n = 111, 58.42\%) & A video that expresses fear, anxiety, worry, or sadness for oneself or others.\\
 & Other (n = 47, 24.74\%) & \\
 & $\bullet$ \textit{Humor (n = 37, 19.47\%)} & A video that presents an unexpected, funny, or playful situation about harassment that is intended to make people laugh.\\
 & $\bullet$ \textit{Hope (n = 4, 2.11\%)} & A video that shows something positive about harassment is happening or will happen, such as decreased threats, increased \\ & & access to support, and/or improved effectiveness of interventions.\\
 & $\bullet$ \textit{Other (n = 6, 3.16\%)} & A type of emotion that cannot be classified into one of the other emotions categories.\\
 & Normal Life Impacts (n = 38, 20\%) & A video that presents how harassment impacts people's normal lives, such as feeling constantly on edge even when harassment \\ & & is not actively occurring.\\
 & Response Efficacy (n = 32, 16.84\%) & A video that demonstrates that precautionary measures are effective in preventing and responding to harassment.\\
 & News (n = 12, 6.32\%) & A video that describes incidents of harassment in a news article.\\
\bottomrule
\end{tabular}%
}
\caption{Codebook categories, example frequencies, and code definitions (continued). * indicates multiple codes could apply. Italicized codes are subcodes and are listed under their primary code.}
\label{tab:codebook_flat}
\end{table}

Since our data included video as well as text data, we incorporated categories established in prior online harassment research \cite{rezvan_analyzing_2020, vogels_state_2021} as well as categories tailored to capturing the nuances of video data. As a result, coding categories were developed inductively, starting with the forms of harassment identified in prior research \cite{Bucchianeri2014multiple, rezvan_analyzing_2020, rezvan2020analyzing, OConnell2000sexual, Bucchianeri2016youth}. 
We assumed that comments depend on the types of the harassment reported in a given video and thus shape the linguistic characteristics and interpretation of the abuse \cite{rezvan_analyzing_2020}. By integrating established macro--domains (Sexual, Physical), critically specific behaviors (Catcalling, Gaslighting), and entities mentioned (Law Enforcement), our codebook balanced analytical specificity with comprehensively describing the observed harassment types.

\subsection{Psycholinguistic Analysis of the Video Comments}
\label{sec: psycholinguistic_analysis_of_videos}

 The 190 videos were uploaded by 129 creators and contained 66,337 comments. We used LIWC to analyze the linguistic qualities of both individual comments and as aggregated video--level texts \cite{tausczik2010psychological}.
 We computed descriptive statistics for each video's likes, views, shares, saves, and categorical metadata (harassment type, location, target, video type, theme, and emotions conveyed).
We assessed the distribution of numeric variables (e.g., LIWC outputs) for skewness and kurtosis, finding most of the numeric variables (97/138) to be moderately to extremely positively skewed and a slightly smaller proportion (94/138) to be leptokurtic. Because the primary harassment categories and subcodes represent nominal classifications rather than continuous measurements, the relationships of interest involved associations between categorical variables. Such data do not satisfy the assumptions required for parametric statistical tests, which assume either normality and continuous or interval--level inputs \cite{curran1996robustness}. Accordingly, $\chi^2$-tests of independence were used to evaluate whether the distribution of contextual variables differed across harassment types \cite{mchugh2013chi}. Bonferroni correction was applied to account for the number of pairwise comparisons and to reduce the risk of Type I error. Where row counts allowed, we also computed Cramer's V \cite{akoglu2018user} to quantify the strength of observed associations, as it provides an interpretable effect size measure specifically suited to $\chi^2$ analyses of categorical data.

To further assess how comments on harassment types differed linguistically, we first examined relationships among all numeric variables using Pearson's correlations \cite{schober2018correlation}. Although 45 variable pairs showed correlations of magnitude $\geq 0.7$, none represented substantively meaningful relationships, as all reflected expected dependencies within LIWC category hierarchies (e.g., between summary variables and their component subcategories) or between core engagement metrics such as views and comments. Correlations involving the harassment type labels yielded only weak negative associations (magnitudes $\leq 0.208$), indicating minimal linear relationships between harassment categories and the numeric features. Mutual information analysis similarly returned 16 variable pairs with scores $\geq 0.6$, but again these involved nested LIWC constructs and therefore did not provide practically meaningful insights. These preliminary analyses confirmed that linear associations were limited and that more appropriate group--comparison methods were needed to assess differences across harassment types.

Given that the LIWC outputs were numeric but exhibited substantial skewness and kurtosis, and because harassment types form multiple independent groups, we used a Kruskal--Wallis test \cite{mckight2010kruskal} ($\alpha = 0.05$) to examine whether the distributions of linguistic features differed across harassment categories. This non--parametric test is well suited for comparing multiple groups when normality assumptions are violated. The test identified 57 LIWC variables that varied significantly across harassment types. We then conducted Dunn's post--hoc tests ($\alpha = 0.05$) to determine which specific harassment type pairs differed, as Dunn's procedure provides an appropriate multiple--comparison adjustment in nonparametric settings. This analysis yielded 32 unique differentiating variables, including 11 LIWC categories that distinguished 12 harassment-type pairs (e.g., sexual harassment vs. bullying differed in the use of anxiety--related language). For each pair, we calculated Cohen's d \cite{cohen1988some} to characterize the magnitude of these differences. The resulting effect sizes were large ($\geq 0.6$) or extremely large ($\geq 1.3$), indicating that these linguistic markers captured substantial and meaningful distinctions among harassment types. Our application of these statistical methodologies to psycholinguistic data are broadly similar to those of \cite{schoenebeck2021drawing, goel2020understanding}.


We examined the linguistic details of each of the 11 LIWC outputs that split the 12 harassment type pairs. For each of the pairs, we performed the following analysis:

\begin{itemize}
    \item Calculated the average score of the differentiating LIWC category for each harassment type (e.g., \texttt{LIWC\_emo\_anx} for sexual vs. bullying).
    \item Identified the most common video type, harassment location, and harassment target for the higher--scoring harassment type.
    \item Selected the top 10 highest--scoring comments per differentiating LIWC category, ensuring they came from distinct videos; for categories with fewer than 10 videos, all relevant top--scoring comments were included.
    \item Compiled the 50 highest--scoring comments overall to examine characteristic linguistic features within each harassment type.
    \item For the stalking--physical harassment pair, where the differentiating category was the composite \texttt{Linguistic} metric, we analyzed both the summary score and each of its 19 subcategories.
    \item Calculated the relative use of all emotion--related LIWC categories and subcategories of \texttt{Linguistic} across harassment types.
\end{itemize}

Importantly, no causal inferences can be drawn from these  differences alone \cite{pearl2009causality}. Our analysis is sufficient to demonstrate statistically and practically significant patterns in the data, but establishing causality would require a greater understanding of the motivations, desires, and goals of video posters and commenters than can be gleaned from this dataset \cite{boyd2012critical}.

\subsection{Stance Detection in the Comments}
\label{sec:methods_stance}

A stratified sample of 100 comments, with harassment types approximately proportional to the full dataset, was manually coded for stance to capture whether comments expressed support, opposition, or neutrality regarding the associated post's author's perspective. Here, stance refers to whether a comment supported, opposed, or remained neutral or ambivalent toward the perspective expressed by the author of the associated post, regardless of the factual status of that perspective \cite{kucuk2020stance, burnham2025stance}. Agreement between human annotation and GPT--4.1--mini classification yielded Cohen’s $\kappa = 0.77$, indicating substantial reliability and comparable performance to prior stance detection studies \cite{landis1977measurement, li2024advancing, liyanage2024gpt}. Consistent with recent work validating large language models (LLMs) as annotation tools \cite{gilardi2023chatgpt, tornberg2025large}, this level of agreement was deemed sufficient to support automated stance labeling for the full dataset. We then assessed the stances of the 66,337 comments across 190 videos using GPT--4.1--mini for few--shot, target--specific stance detection via prompt--based classifications (see Appendix for a more detailed explanation of the model setup) \cite{Pangtey2026-cf, Zhang2024-hh}. Each comment was assigned one of three categories: support, oppose, or neutral. We then computed the overall stance proportions across all comments and constructed \textit{stance x harassment type} contingency tables, allowing for summarization of stance distributions both visually and numerically. To learn preliminary indicators of associations between stance and harassment types, we computed $\chi^2$ tests of independence. While these initial tests did suggest nontrivial associations, the non--independence of the comments within videos and the multi--label observations rendered all $\chi^2$ assessments strictly exploratory.

To accommodate the non--independent and potentially correlated nature of the data, we modeled comment--level stance using a \textit{Bayesian hierarchical multinomial logistic regression} model, including random intercepts for videos to capture clustering and shared contexts and multi--membership harassment type predictors \cite{gelman2007data}. This included estimated posterior stance probabilities for each harassment type and utilized credible intervals to summarize uncertainty. Hierarchical modeling accounts for clustering of comments within videos and allows stance probabilities to vary across videos, providing more reliable inference than models that assume independent observations. Under this model, we defined \textit{polarization} as $Prob(support) + Prob(oppose)$, aligning with distributional conceptions of polarization as movement away from neutrality or concentration in non--neutral positions \cite{dimaggio1996have, Fiorina2008-xr}, and explicitly accounted for video--level contextual variation via random effect terms.

By connecting our stance--based analysis alongside the prior LIWC analyses, we sought to better understand not only whether commenters supported or opposed the views expressed in each video but also the manner in which these stances were expressed. To do this, we restricted analysis to engaged comments (e.g., support vs. oppose) and fit a generalized estimating equation (GEE) logistic model to predict support rate \cite{hanley2003statistical, ziegler2010generalized, liang1986longitudinal}. This model, which accounted for clustering within videos, estimated the independent effects of emotional, power/agency, cognitive/analytic, social, and moral language, allowing us to interpret LIWC as explaining how stance was expressed rather than whether a particular stance occurred. 

%% file: sections/4a-new_results.tex
\section{Results}

Section 4.1 describes the harassment types represented in the dataset and their contextual characteristics. Section 4.2 examines psycholinguistic differences in comments across harassment types using LIWC analysis. Section 4.3 analyzes the stance expressed in comments responding to harassment content. A full accounting of the calculated metrics and statistical results is available in the Appendix.

\subsection{Harassment Types and Contextual Characteristics in the Videos (RQ1)}

This section describes the types of harassment represented in the dataset and the contextual characteristics of the videos in which they appear.

\subsubsection{Harassment Types on TikTok}
\label{sec:harassment_types}
To provide qualitative context for the harassment categories, we briefly describe the types of harassment represented in the dataset and the kinds of interactions depicted in these videos.

\textbf{Sexual harassment} was the most prevalent category in the dataset (n = 94, 49.47\%). These videos frequently depicted inappropriate sexual behavior occurring in public or interpersonal settings, often directed toward women. For example, one video showed a woman responding to a livestream of a man making sexually explicit gestures outside a car occupied by two women. Other videos included commentary or satire about sexual harassment, such as animations mocking women's claims of harassment or comedians joking about experiences with unwanted advances. In many cases, creators framed these incidents through personal narratives or reactions to widely recognizable situations, blending firsthand accounts with commentary intended to raise awareness or spark discussion.

\textbf{Bullying} was the second most common harassment category (n = 44, 23.16\%) and encompassed a broad range of interpersonal conflicts. Some videos documented direct acts of intimidation or aggression, such as footage of a man damaging his ex--wife's property following their divorce. Others involved creators discussing ongoing conflicts with neighbors, coworkers, or acquaintances. In a few instances, videos addressed the legal or social boundaries between bullying and other forms of harassment, such as workplace mistreatment that may be harmful but not legally actionable.

\textbf{Verbal harassment} (n = 37, 19.47\%) typically involved direct insults, threats, or hostile interactions between individuals. Many videos centered on creators responding to abusive comments or recounting confrontations they had experienced offline or online. For instance, one video featured a TikTok user describing how another account repeatedly spammed her comment section with abusive messages, while another showed a creator responding angrily to misogynistic remarks directed at her. In limited cases, creators reenacted or simulated harassment scenarios to illustrate recognizable patterns of verbal abuse.

\textbf{Catcalling} videos (n = 26, 13.68\%) represented a specific subset of sexual harassment in which individuals---typically women---recounted or responded to unsolicited sexual remarks made in public settings. Some videos documented direct encounters, such as male students making crude sexual comments toward female classmates, while others featured creators reflecting on the prevalence of catcalling or expressing solidarity with those who had experienced it. These videos often framed catcalling as a common but frequently dismissed form of harassment, with creators using humor, commentary, or educational messaging to emphasize its broader social impact.

\textbf{Stalking harassment} (n = 25, 13.16\%) included both online and offline monitoring behaviors. Videos in this category ranged from creators documenting individuals who repeatedly tracked or contacted them online to recordings of in--person confrontations. One example showed a creator displaying a list of usernames associated with people allegedly stalking her online, while another depicted a confrontation between a civilian and border patrol officers accused of repeatedly appearing at the individual’s property. These videos often emphasized themes of intrusion, surveillance, and violations of personal space.

\textbf{Law enforcement harassment} (n = 15, 7.89\%) involved encounters between civilians and police officers or other authorities that video creators interpreted as unjustified or abusive. For example, one video showed a police officer threatening a driver during a traffic stop, while another depicted officers questioning two young men about ownership of the vehicle they were driving. These interactions were often recorded from the creator’s perspective and framed as evidence of perceived misuse of authority.

\textbf{Physical harassment} appeared in a small number of videos (n = 10, 5.26\%) and depicted direct physical aggression or violence. Examples included footage of peer aggression as well as creators recounting incidents of assault or unwanted physical contact. One video featured a creator describing being bitten after rejecting a man’s advances, while another used an animated clip to illustrate a scenario in which a girl was deliberately soaked by classmates. These videos frequently emphasized the seriousness and immediacy of physical harm.

\textbf{Gaslighting} appeared relatively infrequently in the dataset (n = 7, 3.68\%), capturing situations in which individuals described being manipulated into doubting their own experiences or perceptions. For example, one video showed a creator discussing the removal of her content from the platform and speculating that her family members had falsely reported her posts, framing the experience as an attempt to silence or undermine her perspective. These videos often foregrounded emotional confusion, frustration, or disbelief, highlighting the psychological dimensions of harassment rather than overtly aggressive behaviors.

Finally, a small set of videos (n = 10, 5.26\%) fell into an \textbf{“other” harassment} category due to the heterogeneous nature of the incidents depicted. These included situations such as individuals impersonating service workers to gain entry to someone’s home, disputes involving online fandom harassment, and videos debating whether certain behaviors constituted harassment at all. The diversity of scenarios in this category reflects the ambiguous boundaries that sometimes surround harassment discourse on social media platforms.

\subsubsection{Contextual Characteristics of Harassment Types on TikTok}
\label{sec:context_details}
Dominant contextual characteristics for each harassment type are depicted in Table~\ref{tab:harassment_characteristics}.

Across the dataset, harassment was most frequently depicted as an interpersonal experience occurring in face--to--face contexts. In--person encounters dominated nearly all harassment categories, with relatively few videos depicting exclusively online harassment or scripted scenarios. Creators also commonly framed harassment through firsthand experiences, with personal narratives emerging as the most prevalent thematic approach across harassment types.

Patterns of harassment targets varied by category. Bullying, verbal harassment, gaslighting, stalking, and law enforcement harassment most often centered on the uploader, reflecting firsthand documentation or personal testimony. In contrast, sexual harassment, catcalling, physical harassment, and videos categorized as “other” more frequently focused on incidents involving third parties. Celebrity targets were rare overall but appeared slightly more often in physical harassment and bullying--related content. Beyond merely indicating the relationship between uploader and target, these findings suggest the presence of distinct communities within the dataset: some harassment types are more often characterized by self--reported incidents, while others more frequently center on harassment reported on behalf of someone other than the uploader. 

\begin{table}[t]
\centering
\resizebox{\textwidth}{!}{%
\begin{tabular}{lcccccc}
\toprule
\textbf{Harassment Type} & \textbf{N (\%)} & \textbf{Typical Target} & \textbf{Typical Location} & \textbf{Most Common Video Type} & \textbf{Dominant Theme} & \textbf{Dominant Emotion} \\
\midrule
Sexual & 94 (49.47\%) & Third party (60.20\%) & In person (85.71\%) & Oral speech (38.78\%) & Personal experience (35.85\%) & Alarm/concern (37.69\%) \\
Bullying & 44 (23.16\%) & Uploader (61.36\%) & In person (68.18\%) & No dominant category & Personal experience (58.33\%) & Alarm/concern (57.41\%) \\
Verbal & 37 (19.47\%) & Uploader (70.27\%) & In person (72.97\%) & Oral speech (37.84\%) & Personal experience (79.49\%) & Alarm/concern (56.82\%) \\
Catcalling & 26 (13.68\%) & Third party (76.92\%) & In person (92.31\%) & No dominant category & Knowledge/awareness (62.96\%) & No dominant category \\
Stalking & 25 (13.16\%) & Uploader (60.87\%) & In person (91.30\%) & No dominant category & Personal experience (58.33\%) & Alarm/concern (56.76\%) \\
Law enforcement & 15 (7.89\%) & Uploader (86.67\%) & In person (100\%) & POV (60\%) & Personal experience (80\%) & Alarm/concern (65\%) \\
Physical & 10 (5.26\%) & Third party (60\%) & In person (90\%) & No dominant category & No dominant category & Alarm/concern (63.64\%) \\
Other & 10 (5.26\%) & Third party (70\%) & Other (40\%) & Other (40\%) & Other (54.55\%) & Other (36.36\%) \\
Gaslighting & 7 (3.68\%) & Uploader (71.43\%) & In person (85.71\%) & Talking over background image (42.86\%) & Personal experience (71.43\%) & Alarm/concern (75\%) \\
\bottomrule
\end{tabular}
}
\caption{Dominant characteristics represent the most frequently observed category within each harassment type. Categories were considered dominant when they accounted for at least 35\% of observations; otherwise, the characteristic was labeled as having no dominant category. All percentages are rounded to 2 significant decimal places.}
\label{tab:harassment_characteristics}
\end{table}

Video formats were more diverse than other contextual variables. The most common format was oral speech, in which creators spoke directly to the camera to recount harassment incidents or express their reactions. This format emphasized personal testimony, allowing creators to communicate emotion through facial expressions, tone of voice, and direct address. Point-of-view (POV) recordings frequently documented interactions as they unfolded from the creator's perspective, often functioning as visual evidence of harassment by showing the surrounding context rather than the creator themselves. While oral speech and POV recordings were common across several harassment types, creators employed a wide range of communicative strategies, including responses to stitches, commentary over background imagery, and simulated reenactments. 
Platform-native features such as stitches enabled creators to directly respond to harassing content by embedding the original video alongside their commentary, creating a dialogic form of evidence and rebuttal.
These formats help describe the harassment targeting individuals other than the uploader, as stitch responses and reenactments inherently rely on externally referenced or reconstructed events. Some creators also narrated over screenshots, photographs, or other background imagery to provide contextual information or supporting evidence, while others adopted informational formats to explain broader patterns of harassment.
This diversity suggests that harassment experiences on TikTok are communicated through multiple narrative and stylistic conventions rather than a single dominant format.

Emotionally, harassment--related videos were overwhelmingly framed through alarm or concern, reflecting the seriousness with which these experiences were presented. Humor and response efficacy appeared less frequently but remained present across several harassment categories, indicating that creators sometimes combined serious messaging with coping strategies or ironic commentary.

\subsubsection{Descriptive Statistics of Videos and Comments per Harassment Types and Engagement Characteristics}
\label{sec:data-descriptives}
Engagement levels varied substantially across videos (Table~\ref{tab:harassment_summary}), with average comment counts typically exceeding medians, reflecting the skewed distribution common to social media engagement metrics. More than half of the videos in most harassment categories received over 100 comments, indicating that harassment--related content in our dataset often generated substantial audience interaction. 

\begin{table}[h]
\centering
\resizebox{\textwidth}{!}{%
\begin{tabular}{lrrrrrrrrr}
\toprule
\textbf{Harassment Type} & \textbf{Sexual} & \textbf{Bullying} & \textbf{Verbal} & \textbf{Catcalling} & \textbf{Stalking} & \textbf{Law Enf.} & \textbf{Physical} & \textbf{Other} & \textbf{Gaslighting} \\
\midrule
Number of Videos & 94 & 44 & 37 & 26 & 25 & 17 & 10 & 9 & 8 \\
Avg. Comment Word Count & 12.45 & 10.92 & 11.22 & 13.73 & 11.92 & 11.25 & 10.46 & 11.55 & 10.93 \\
Avg. Number of Comments & 335.06 & 364.25 & 523.14 & 260.42 & 554.36 & 473.29 & 481.50 & 182.56 & 191.50 \\
Median Number of Comments & 178 & 199 & 409 & 159 & 425 & 207 & 370 & 107 & 80 \\
Avg. Number of Likes & 43,452 & 32,617 & 89,953 & 23,992 & 64,563 & 21,955 & 83,962 & 88,339 & 7,397 \\
Avg. Number of Views & 451,567 & 493,277 & 1,048,395 & 128,108 & 864,488 & 398,813 & 926,540 & 946,382 & 117,875 \\
Avg. Number of Shares & 682.71 & 395.09 & 876.03 & 515.19 & 510.88 & 387.93 & 1731.3 & 725.91 & 52.25 \\
\bottomrule
\end{tabular}%
}
\caption{Summary Statistics by Harassment Type}
\label{tab:harassment_summary}
\end{table}

\begin{figure}[H]
\centering
\includegraphics[width=\textwidth]{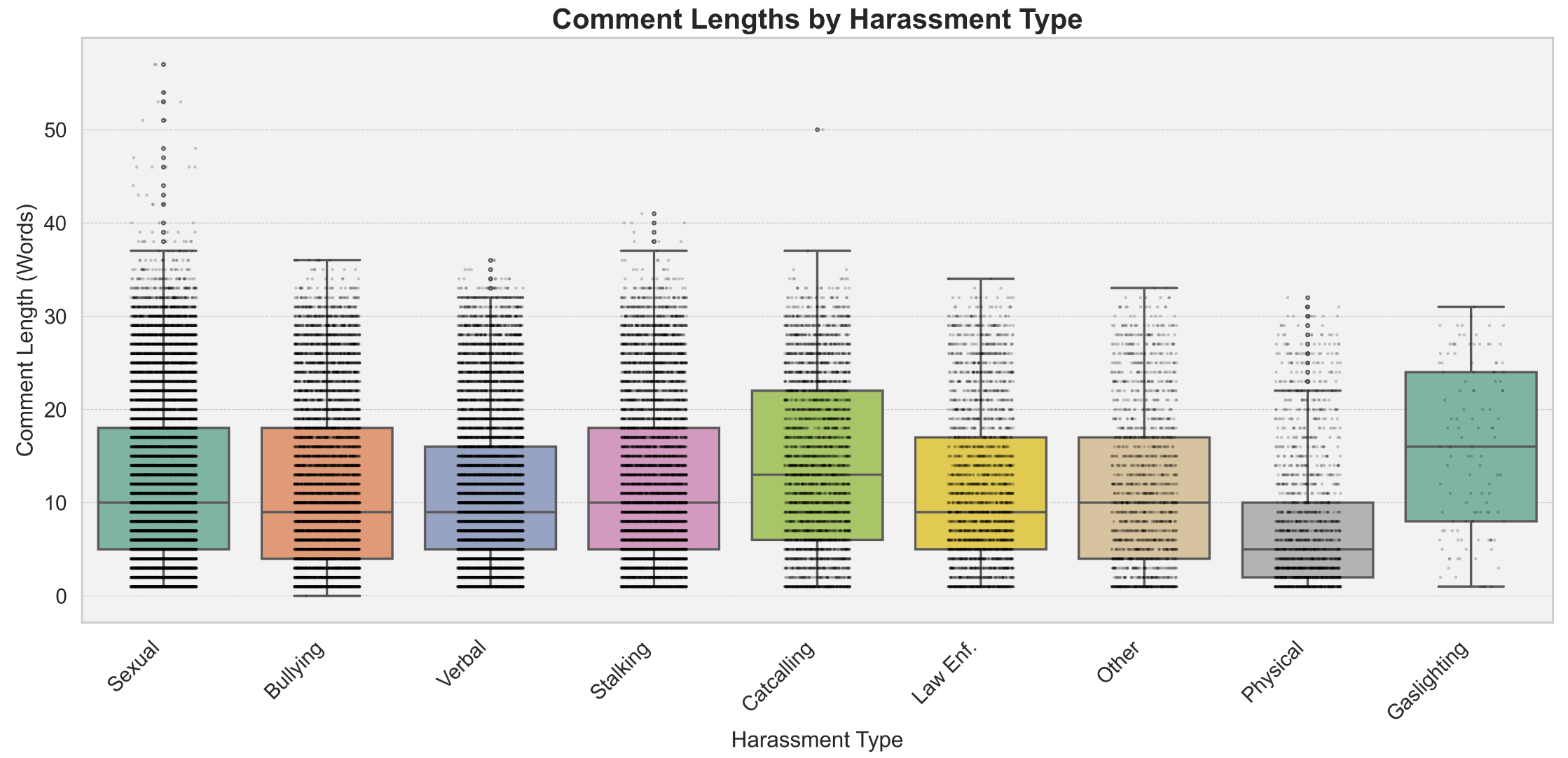}
\caption{Comment Lengths by Harassment Type. Each box ranges from the 25th to 75th percentile, with the middle line representing the median comment length. Points outside the whiskers represent outliers. Darker horizontal lines indicate a greater number of comments with the corresponding word count.}
\label{fig:comment_lengths}
\end{figure}

Statistical analyses further indicated differences in engagement patterns across harassment categories. A significant difference in view counts was observed between catcalling and stalking videos ($p = 0.0284$), with stalking videos typically receiving substantially higher view counts. Average comment length differed significantly across harassment types (\textit{Kruskal-Wallis} $p < 0.001$), with sexual and physical harassment videos exhibiting greater variability and more extreme outliers in comment length (Figure~\ref{fig:comment_lengths}). This implies that comments responding to sexual and physical harassment videos were less uniform in length than those responding to other harassment types, suggesting that the depth of commenter engagement with the videos’ content may be less predictable.
The Kruskal--Wallis test also indicated a significant overall difference in the number of comments across harassment types ($H = 14.1158$, $p = 0.0492$), though no pairwise comparisons remained significant after correction. No significant differences were observed in the number of likes ($H = 13.1294$, $p = 0.69$) or shares ($H = 7.7249$, $p = 0.3575$). 

\begin{table}[h]
\centering
\resizebox{0.7\textwidth}{!}{%
\begin{tabular}{c|c|c}
\toprule
\textbf{Number of Harassment Types} & \textbf{Mean Comments Per Video} & \textbf{Mean Likes Per Video} \\
\midrule
1 & 249 & 23,238 \\
2 & 480 & 75,237 \\
3 & 584 & 44,807 \\
\bottomrule
\end{tabular}%
}
\caption{Engagement by Number of Harassment Type Labels}
\label{tab:harassment_label_distribution}
\end{table}

Notably, 79 videos (41.58\%) received multiple harassment labels, reflecting the presence of overlapping or multi--faceted harassment experiences within a single video. Videos assigned multiple harassment labels were also associated with higher engagement metrics, including increased average comment counts and likes (Table~\ref{tab:harassment_label_distribution}), suggesting a positive association between multi-label narratives and viewer engagement.

\subsection{LIWC--Differentiated Harassment Types in the Comments (RQ2)}
\label{resultsRQ2}

A total of 12 harassment type pairs were distinguished based on statistically significant differences in the LIWC scores of their associated comments. Although all variable pairs were assessed for statistical significance, many differences lacked practical significance due to redundancy or hierarchical overlap within LIWC categories. Accordingly, we report only those differences that were both statistically and practically meaningful, where practical significance was signaled by large or extremely large effect sizes (Table~\ref{tab:liwc_kruskal_dunn}). 

\begin{table}[H]
\centering
\resizebox{\textwidth}{!}{%
\begin{tabular}{llrrrrl}
\toprule
\textbf{Harassment Type 1} & \textbf{Harassment Type 2} & \textbf{Kruskal-Wallis P-Value} & \textbf{Dunn Score} & \textbf{LIWC Category} & \textbf{Cohen's d} & \textbf{Effect Size} \\
\midrule
Law Enforcement & Bullying & 0.004075 & 0.0001320 & Culture & 1.42478 & Extremely Large \\
Other & Bullying & 0.013486 & 0.0003746 & number & 1.55888 & Extremely Large \\
Sexual & Bullying & 0.003353 & 0.0000931 & emo\_anx & 0.60879 & Large \\
Stalking & Bullying & 0.000026 & 0.0000007 & emo\_anx & 1.57032 & Extremely Large \\
Verbal & Bullying & 0.046131 & 0.0012814 & visual & 0.55909 & Large \\
Sexual & Physical & 0.038093 & 0.0010581 & adj & 1.60731 & Extremely Large \\
Verbal & Sexual & 0.024711 & 0.0006864 & Clout & 0.77021 & Large \\
Stalking & Sexual & 0.046611 & 0.0012947 & home & 1.21642 & Extremely Large \\
Catcalling & Stalking & 0.014922 & 0.0004145 & emo\_pos & 1.54803 & Extremely Large \\
Stalking & Physical & 0.046750 & 0.0012984 & Linguistic & 3.14054 & Extremely Large \\
Physical & Catcalling & 0.023428 & 0.0006508 & QMark & 3.12215 & Extremely Large \\
Law Enforcement & Verbal & 0.041072 & 0.0011409 & space & 1.03308 & Very Large \\
\bottomrule
\end{tabular}%
}
\caption{Comparisons of harassment types with LIWC category differentiation and effect sizes. Categories in "Harassment Type 1" column had higher average scores than those in "Harassment Type 2" column. Some of LIWC categories are aggregate categories (e.g., \texttt{Linguistic}) and are composed of multiple subcategories. While every harassment type pair was tested across all LIWC categories, some of the significant effects appeared only for the aggregate categories, implying that the subcategories did not independently differentiated the harassment type pair(s). For a full breakdown of the subcategories of \texttt{Linguistic}, see Table~\ref{tab:liwc_linguistic_stalking_physical} in the Appendix.}
\label{tab:liwc_kruskal_dunn}
\end{table}

Across harassment types, several recurring linguistic patterns emerged. The most pronounced differences, based on Cohen's d effect size scores, appeared in comments responding to physical harassment videos. These comments showed reduced use of function words, pronouns, verbs, and adjectives, alongside increased use of number and quantity words, as exhibited by comments such as\footnote{In all subsequent quotations in Section~\ref{resultsRQ2}, we have underlined the words associated with the relevant, elevated LIWC categories.}:

\begin{quote}
    \textit{This is so useful for the disabled community...We’re \underline{4x} more likely to be abused- I want to be \underline{4x} more prepared then they expect.}
\end{quote}

This pattern suggests a comparatively concrete and pragmatic mode of expression, with commenters emphasizing factual or collective details rather than elaborative narrative descriptions. The focus on groups over individuals also helps to explain the prominence of number and quantity--related words, as such details were often included to contextualize a commenter's remarks. 

In contrast, comments on catcalling videos exhibited elevated use of positive emotion words, function words, and first--person singular pronouns. As shown by comments such as:

\begin{quote}
    \textit{\underline{I} \underline{love} this so much. So many \underline{amazing} points made. Thank you for standing up for all of us lonely ladies}
\end{quote}

\noindent catcalling videos were often met with supportive and positively--phrased comments. This linguistic profile reflects a more emotionally expressive and self--referential style of engagement, consistent with comments that frequently expressed solidarity, affirmation, or personal identification  with the experiences described in the videos.

Comments responding to law enforcement harassment displayed a different linguistic profile. These comments showed decreased use of function words and first--person singular pronouns but increased use of second-- and third--person plural pronouns. This indicates a greater focus on collective actors or institutional dynamics, suggesting that commenters often framed these encounters in terms of broader social groups or systemic behavior rather than individual experiences. As demonstrated by comments like:

\begin{quote}
    \textit{If \underline{he} attempts another stop safely call 911 and explain \underline{you} are being targeted and driving to the nearest police station because \underline{you} feel unsafe.}
\end{quote}

Many comments responding to law enforcement harassment videos emphasized advice--giving or other--directing suggestions, frequently instructing others on how to respond. While not always the case, many of these comments also attributed qualities such as ‘crazy’ or ‘creepy’ to perpetrators, a pattern also observed in comments on stalking harassment videos.

Several other harassment types---including gaslighting, bullying, and inappropriate sexual behavior---showed more moderate but still meaningful variation across LIWC categories. Differences within these categories were most often associated with emotional language (e.g., anxiety--related terms), pronoun use, and stylistic markers such as clout or visual language. Together, these findings indicate that harassment discourse on TikTok is not linguistically uniform; instead, commenters adopt distinct rhetorical and emotional styles depending on the type of harassment being discussed. 

The use of first--person and emotional language (e.g., sexual harassment), alongside third--person plural language with institutional framing (e.g., law enforcement), similar to the findings about harassment targets (see Section~\ref{sec:context_details}), further suggests the presence of distinct groups within harassment--related discourse. First--person and emotional language foregrounds personal narrative tones, while third--person plural pronouns emphasize a collective, group--level interpretation of the events described in a given video.

\subsection{Examining the Stances of the Comments (RQ3)}
\label{sec:results_stance}

Comments were predominantly neutral (approx. 48\%), suggesting that nearly half of all comments expressed neither clear support nor opposition to the associated video's content. When commenters did adopt a non--neutral stance, however, they were substantially more likely to express support rather than opposition. This asymmetry persisted even when the analysis was restricted to only non--neutral comments, where supportive responses constituted approximately 71\% of polarized discourse.

That said, stance distributions did differ across harassment types. Some types---bullying (42.69\%) and stalking (57.12\%)---were more likely to elicit non--neutral responses, while others---physical (74.75\%) and “other” (69.92\%) harassment---were associated with elevated neutrality (Appendix~\ref{tab:stance_residuals}). This implies that the type of harassment shaped not only how commenters reacted but whether they reacted at all.

The hierarchical model showed that \textit{polarization}—the probability of a non--neutral stance—varied across harassment types, though most differences were relatively small (Appendix~\ref{tab:polarization_contrasts}). Only bullying clearly exceeded law enforcement harassment (+0.17 mean polarization difference) and “other” harassment (+0.16 mean polarization difference) in polarization, while many other harassment type differences were indistinguishable once uncertainty resulting from video--level variation was accounted for (Appendix~\ref{tab:posterior_stance_probs}). Put differently, while harassment type influenced the degree of polarization for a given video, only a few harassment types substantially increased non--neutral engagement.

Across all harassment types, increases in polarization were largely driven by supportive responses rather than opposition; oppositional comments remained consistently less frequent. Notably, there was also significant video--level variation in stance. Within a single harassment type, different videos generated substantially different levels of engagement ($\sigma_{\text{video}} \approx 1$ (log-odds)), indicating that contextual factors such as framing, tone, and audience nontrivially influenced how commenters responded to videos (Appendix~\ref{tab:video_random_effects}): the \textit{specific video} often mattered as much as the harassment type.

Among non--neutral comments, linguistic style strongly differentiated supportive from oppositional engagement. Comments employing more emotional and power or agency--related language were more likely to be supportive, whereas those with greater use of cognitive or analytic language were more likely to be oppositional (Appendix~\ref{tab:gee_liwc}). The use of social and moral language did not independently predict stance. Accordingly, these results suggest that supportive engagement was affective and empowering, whereas oppositional engagement was analytic and deliberative. Finally, the modeling showed that stance was not reducible to sentiment. Instead, stance reflected positional alignment, while linguistic features captured how that position was expressed, revealing distinct affective versus analytic patterns within polarized discourse.

%% file: sections/5-Discussion.tex
\section{Discussion}
\subsection{Implications of Harassment Types and Characteristics (RQ1)}



While our results confirm previous findings showing sexual harassment to be the predominant type of online harassment \cite{henry2018technology}, 
TikTok--specific practices akin to those described in sociotechnical harm taxonomies by Shelby et al. \cite{shelby2023socio} and Ugwudike et al. \cite{Ugwudike2024} were also salient. 
In particular, forms of risk--producing visibility and technology--facilitated violence \cite{shelby2023socio} were readily apparent in the dataset. Here, however, such practices manifest through video creators documenting harassment on behalf of others, often without clear evidence of explicit consent. Even if these creators were wholly well--intentioned, an assumption for which we have little reason to doubt, the publicizing of digital harassment introduces nontrivial risks to those targeted: it is well--established that online aggression can and regularly does translate into offline violence \cite{donato2022offline, Trottier2021-cv}.

Although most videos documented 
harassment experienced by the uploader, a substantial share focused on incidents involving others. This suggests that harassment--related content on TikTok is not monolithic and may serve multiple communicative purposes, including personal disclosure, awareness raising, and commentary on broader social experiences. The prominence of both self--reported incidents and reports made on behalf of others highlights that, similar to prior accounts of offline harassment \cite{thornton2012prevalence, jones2015victim}, the person who ultimately reports harassment varies depending on the nature of the harassment. In particular, harassment that is inherently tied to physical violence is less likely to be reported by the victim \cite{thornton2012prevalence}, an unfortunate reality reflected in our dataset. 
Nevertheless, as noted by \cite{kim2021human, Razi2021KIm}, neither linguistic nor statistical analysis alone fully captures how harassment manifests online: qualitative context remains essential for understanding these dynamics.

Notably, sexual harassment--focused videos did not attract the greatest levels of user engagement. On average, verbal harassment videos received the most visibility, as they were the only category with average view counts exceeding one million. As Milton notes \cite{milton2023see}, TikTok’s short--form content facilitates the rapid spread of information across communities, allowing certain narratives to gain traction through engagement. 
Thus, prominence of a harassment type does not necessarily imply ethical acceptance, but may instead reflect creator incentives to feature it.

Given the high turnover of popular creators on TikTok \cite{Bleier2024} and the opaque nature of its content recommendation algorithm \cite{klug2021trick}, creators often optimize videos around visible engagement metrics. Consequently, harassment types that attract fewer views may receive less attention from creators. For example, gaslighting videos typically received the lowest view counts in our dataset, suggesting lower engagement potential and, consequently, weaker incentives for creators to address it. 
TikTok’s average video length of roughly 35 seconds \cite{socialinsider_tiktok_videos_length} further reinforces a preference for content that captures attention quickly, often at the expense of depth or nuance \cite{molem2024keeping}. 
The impact of a given video is therefore strongly shaped by its visibility \cite{OBADIMU2021102660}. As a consequence, the relationship between what becomes visible and how users engage with it is nontrivial. Decisions about which content is promoted---whether by moderators, algorithms, or users themselves---affect both the emotive responses of users \cite{taylor2019accountability} and their likelihood of proactively addressing harassment \cite{wong2021standing}. That said, engagement metrics alone appear insufficient for distinguishing harassment types with the depth and accuracy required for automated detection or integration into moderation systems. 
Instead, contextual characteristics---including targets, locations, formats, themes, and emotions---may provide the additional insight required to fully understand the nuance of how harassment is represented on TikTok.


Importantly, contextual features alone do not fully capture how harassment manifests on TikTok. While certain types displayed distinctive contextual patterns, many characteristics overlapped across categories. Harassment experiences on the platform are thus communicated through combinations of narrative framing, platform affordances, and audience engagement practices rather than through a single defining contextual feature. This, in turn, influences how users engage with harassment content and gives rise to distinct patterns of interpretation.
These findings highlight the limitations of treating online harassment as a single, unified phenomenon, an approach still common in many studies \cite{ghosh-chowdhury-etal-2019-speak, khatua2020, liuYing2019}. Our results instead support calls for more context--sensitive approaches to harassment detection and analysis \cite{Razi2021KIm, rheault2019line, marwick2014s}. Moreover, the persistence of harassment--related content despite moderation efforts suggests that contextual features of videos alone are insufficient for distinguishing between harassment, documentation, support, and awareness--raising on TikTok.
For this reason, examining the psycholinguistic characteristics of user comments provides an important complementary perspective.

\subsection{Implications of Psycholinguistic Differences in Harassment (RQ2)}
Our results suggest that the psycholinguistic characteristics of comments vary meaningfully across different forms of harassment and may provide insight into how audiences interpret and respond to harassment--related content on TikTok. Rather than exhibiting a single linguistic style, commenters appear to adopt distinct rhetorical and emotional patterns depending on the harassment being discussed, thereby constituting distinct orientations to harassment experiences. 

Contrary to expectations given harassment's harmful nature, several harassment categories elicited comments containing substantial positive emotional language. In particular, sexual harassment videos were associated with increased use of positive emotion vocabulary alongside elevated use of first--person singular pronouns. These patterns suggest that commenters frequently framed their responses through personal reactions or experiences rather than focusing exclusively on the victim. Such self--referential engagement may reflect attempts to express empathy, solidarity, or shared experience \cite{HuangLu2024, Beasley2015emotional}. In this sense, responses to sexual harassment content often centered on how viewers themselves related to the events depicted in the video rather than remaining external observers.

By contrast, comments responding to gaslighting and law enforcement harassment exhibited greater use of third--person plural pronouns, indicating that commenters often framed these events in terms of collective actors or institutional dynamics. 
This aligns with the institutional nature of law enforcement encounters as well as the socially mediated dynamic often associated with gaslighting \cite{Irigaray2023}. Correspondingly, these categories were associated with comparatively lower levels of emotional language. This suggests that commenters may have engaged with these videos through a more analytical or socially contextualized lens rather than through direct emotional identification with the victim, suggesting a more distanced or observational mode of engagement \cite{semin2002cultural}. 
Physical harassment comments displayed a similarly restrained emotional profile. Because physical harassment involves immediate and observable harm, commenters may have focused more heavily on the factual details of the event rather than on emotional interpretations. The reduced use of descriptive and emotional language in these comments may therefore reflect attempts to process the incident as a concrete occurrence rather than as a psychologically interpreted experience. 

Bullying emerged as one of the most linguistically distinct harassment type, differing from several others harassment types across multiple LIWC dimensions. However, the specific linguistic features separating bullying from other forms of harassment varied across comparisons. In some cases---such as the law enforcement--bullying pair---the distinction was marked by the use of culturally oriented language, reflecting the institutional and legal contexts often invoked in discussions of policing.
Other distinctions---such as the separation of bullying from verbal harassment through visual language or from "other" harassment through number--related language---may reflect the situational nature of bullying discourse, which frequently references appearances, social status, or performance metrics \cite{ringrose2020normative}. Similar associations have been well--documented in other domains of online communication \cite{Abitbol2018, Sap2019}, suggesting that comparable linguistic dynamics may also emerge in discussions of harassment on TikTok. 

Differences between bullying, sexual harassment, and stalking were also reflected in the use of anxiety--related language captured by the LIWC's anxious emotions category. As prior research has shown, experiences of sexual harassment and stalking often produce substantial psychological distress \cite{McHugh2018WhenSM, TausczikPennebaker2010}. These forms of harassment may therefore prompt commenters to express concern or anxiety more frequently than in discussions of bullying. Stalking, in particular, is associated with persistent monitoring and perceived loss of personal autonomy, experiences that can produce sustained fear or uncertainty \cite{Mullen1999, Miller2012, surviving_stalking_impact}. These dynamics help explain the elevated use of anxiety--related language observed in comments responding to stalking-related content. Finally, sexual and verbal harassment were distinguished by differences in clout--related language, suggesting that commenters framed these experiences in terms of perceived power dynamics \cite{drouin2017linguistic, mohammadinodooshan2025successful}. Verbal harassment comments more frequently contained language associated with confidence or authority, which may reflect attempts by commenters to evaluate, challenge, or reinterpret the interaction. Because power relations often depend on contextual factors such as the identities of the individuals involved or the setting in which the interaction occurs \cite{BeatonAli2024}, these linguistic differences may also reflect commenters' attempts to interpret the underlying social dynamics of the harassment event. Collectively, these findings indicate that harassment discourse on TikTok is not linguistically uniform. Instead, commenters adopt different emotional, rhetorical, and interpretive strategies depending on the type of harassment depicted. Psycholinguistic features therefore offer a complementary lens through which harassment--related discourse can be understood, highlighting how audiences interpret, evaluate, and respond to harassment experiences on the platform depending on their relationship to the harassment.

\subsection{Implications of Commenters' Stance Toward Harassment Content (RQ3)}

The stance analysis provides insight into how audiences interpret and respond to harassment narratives on TikTok. Across the dataset, comments were most often neutral, suggesting that many users engaged with harassment--related videos without explicitly expressing support or opposition. This pattern may reflect the participatory nature of TikTok’s comment culture, where users frequently contribute short reactions or observations rather than explicitly evaluating the content \cite{medranda2026emotional}. Given the platform’s rapid engagement dynamics and algorithmic amplification of active discussions \cite{TikToktradeoff}, comments may function less as formal judgments and more as conversational participation within ongoing discourse. Accordingly, neutral comments do not necessarily imply that the commenters are themselves neutral about harassment when the topic is given due consideration. Instead, their expressed neutrality may signal the rapid movement through algorithmically--served content common to social media use \cite{firth2019online}, though genuine neutrality remains possible. 

When commenters did adopt a stance, however, supportive responses substantially outnumbered oppositional ones. This pattern aligns with prior work documenting how social media platforms can facilitate community support and collective responses to harmful experiences \cite{andalibi2016,Leopold2019TheHH}. Harassment--related content on TikTok may therefore function, at least in part, as a space where users express validation or solidarity with individuals describing harmful encounters \cite{medranda2026emotional}, particularly when commenters position themselves in relation to the experience being described. Such behavior resembles the dynamics observed in other online contexts where individuals share experiences of harassment or trauma and receive supportive responses from broader communities \cite{levaot2021making}. That said, the presence of neutral and oppositional comments indicates that harassment discourse on TikTok is not uniformly supportive, and that audience reactions may vary depending on how the incident is interpreted or framed. Moreover, variation in stance can reflect differences in how commenters relate to the harassment event \cite{sherif1973personal}, though the connection between personal circumstances and depth of commitment to any particular stance may not be straightforward \cite{crosby1986introduction}. Contrary to what might be assumed, greater similarity between the experiences of the harassed individual and the stance--maker does not necessarily imply stronger attachment to the subject.

Importantly, the hierarchical model indicated substantial variation in stance across individual videos, even within the same harassment category. This finding suggests that the specific narrative framing of a video may shape audience responses as strongly as the underlying harassment type. Prior research has emphasized that individuals interpret harassment narratives through the lens of their own experiences, identities, and relationships to the content \cite{kim2021human}, positioning themselves as either participants or observational evaluators. In the context of TikTok, these interpretive biases may be further influenced by platform affordances such as short video formats, narrative storytelling conventions, and algorithmic recommendation systems that expose users to content beyond their immediate social networks \cite{Milton2023}. As a result, two videos depicting similar forms of harassment may elicit different audience reactions, reflecting not only differences in personal stance but also participation in distinct interpretive communities.

Linguistic differences between supportive and oppositional comments further suggest that stance is expressed through distinct rhetorical modes of engagement, encapsulating different orientations toward the harassment event. Supportive comments were more likely to contain emotional language and expressions associated with agency or empowerment, whereas oppositional comments more frequently contained cognitive and analytic language. This distinction indicates that supportive engagement may emphasize affective identification with the victim or situation, while oppositional engagement may involve more deliberative or evaluative responses to the content \cite{TausczikPennebaker2010, pennebaker2011secret, greene2004neural}. These differences further reinforce the idea that audience responses to harassment narratives are not solely determined by the type of harassment depicted, but also by the interpretive frameworks through which commenters engage with the video. These findings suggest that harassment discourse on TikTok operates as a complex interpretive space where users participate in conversations about harmful experiences through a mixture of supportive, neutral, and critical responses. Understanding these stance dynamics complements the contextual and psycholinguistic analyses presented earlier, highlighting how both video narratives and audience discourse contribute to the broader social meaning of harassment--related content on the platform. 
The current design of TikTok treats videos with the "harassment" hashtag as a "flat" hierarchy (since they are all placed under the same hashtag) but, for the users, they serve different meanings: sharing experiences to be witnessed, raising awareness, and or serving as attention-soliciting content. Future design implications should allow one to further distinguish these different goals and/or provide different algorithmic visibilities to awareness causes, like harassment, where these creators feel safe to share such content. Future studies should engage in potential design applications to support a range of purposes for which these creators shared their content.

%% file: sections/6-Conclusion.tex
\section{Limitations and Future Work}

Due to TikTok's moderation opacity \cite{chang2025opaque} and the lack of public information regarding the reasons for a given video's removal (i.e., whether a user chose to remove it or if it violated content guidelines), we cannot assess why videos included in our dataset were removed. Also, in Section~\ref{sec:data_collect}, we acknowledged the short time window of our dataset and noted that some of the videos used have since been removed from TikTok. That said, differences in the longevity of videos addressing different types of harassment may reveal further insights into the nature of harassment on TikTok. As such, future studies should address the broader question which types of harassment--related videos persist on TikTok. Similarly, longitudinal analyses should be conducted to identify what purposes, beyond labeling videos as related to harassment, the hashtag \#harassment may serve on TikTok. Given that
hashtags are known to serve multiple purposes online \cite{zelenkauskaite2025hashtags}, understanding the potentially multifunctional nature of \#harassment is of nontrivial importance. 

We also acknowledge that TikTok, like many social media platforms, evolves rapidly as its user base, content practices, and recommendation systems change. It is, therefore, possible that some aspects of how harassment is represented and discussed on the platform shift over time. That said, many of the harassment behaviors observed in our dataset closely resemble patterns documented in offline contexts (e.g., sexual harassment \cite{OConnell2000sexual} or gaslighting, \cite{Irigaray2023}). Given the relatively stable nature of many offline harassment behaviors over time \cite{rueger_stability_2011}, it is plausible that the broader dynamics identified in this study remain relevant even as platform practices evolve.

Another limitation of our study stems from the constraints of the API used to retrieve TikTok data. The API provides access only to a finite set of videos surfaced through the platform's recommendation system. As a result, the dataset reflects algorithmically promoted content rather than a random sample of all videos containing the hashtag \#Harassment. Consequently, our analysis captures the types of harassment--related videos that TikTok’s recommendation system makes visible to users, rather than the totality of harassment--related content on the platform. 


Future research should expand upon and operationalize these findings. Re-collecting algorithmically served TikTok videos would enable temporal analyses examining how harassment discourse changes over time. Large--scale data collection could also support the development of classifiers capable of distinguishing among different types of harassment. If sufficiently accurate, such models could be incorporated into existing moderation systems to improve the effectiveness of harassment detection and response.

Our analysis revealed distinguishing features for 24 of the 36 possible harassment type pairs, leaving 12 pairs without clearly differentiating characteristics. Future work may be able to resolve these distinctions by incorporating additional contextual variables, such as video length, creator characteristics, or account metadata. Because many of these details were not accessible through the API used in this study, addressing this limitation will likely require alternative data collection strategies or collaboration with platform providers.

Finally, TikTok videos and their comments constitute a multimodal form of data, with the particular medium used---images, direct-to-camera speech, text-only posts, or some combination thereof---likely playing a significant role in how harassment is interpreted. While our coding of the videos touched on this issue through video type, greater nuance could be gained from a deeper analysis of the posts’ audiovisual designs. Accordingly, future research on harassment on TikTok and other multimodal social media platforms should conduct more granular analyses of medium-specific features to better explain why different forms of harassment may elicit different reactions from commenters.

\section{Conclusion}

Harassment on TikTok is not a monolithic phenomenon but a multifaceted one shaped by platform affordances, narrative context, and audience interpretation. Through a mixed--methods analysis of videos and comment discourse associated with the \#harassment hashtag, this study showcases distinct contextual and psycholinguistic patterns across harassment types and provides insights how audience responses vary depending on how harassment narratives are framed. Thus, these findings highlight the importance of considering both content characteristics and audience reaction discourse when studying online harassment and suggest opportunities for more context--sensitive approaches to harassment detection, moderation, and platform design.


%% file: sections/7a-new_appendix.tex
\section{Appendix}
\label{sec:new_appendix}

\noindent \textbf{Explanation of the GPT--based Annotation:} We utilized GPT-4.1-mini with the following prompts to annotate comment stance:
\begin{itemize}
\item \underline{System Prompt} - You are an expert social media content analyst. Your task is to classify the stance of a comment responding to a TikTok video about harassment. Stance is judged relative to the associated post author's expressed perspective or framing. In other words, determine whether the comment affirms, rejects, or remains ambivalent toward the post author's perspective, regardless of whether the author's perspective is factually correct. You must follow the labeling style shown in the examples exactly.
\item \underline{Task Prompt} - You are labeling comments posted under TikTok videos about harassment. Classify each comment according to its stance toward the associated post author's expressed perspective or framing. Each comment must be labeled as exactly one of:
\begin{itemize}
    \item \texttt{support}: the comment agrees with, affirms, sympathizes with, validates, or otherwise supports the post author's perspective or framing.
    \item \texttt{oppose}: the comment rejects, criticizes, dismisses, mocks, or otherwise challenges the post author's perspective or framing.
    \item \texttt{neutral}: the comment does not clearly support or oppose the post author's perspective, is ambiguous, asks for information, changes the subject, or cannot be confidently classified.
\end{itemize}
First, check the language of the comment. If the comment is not clearly written in English, or is mostly in another language, you MUST label it as: \texttt{neutral}. Do not attempt to infer stance from non-English text.
Your output must be EXACTLY one word: \texttt{support}, \texttt{neutral}, or \texttt{oppose}. Do not add anything else.
Here are examples:
\begin{itemize}
    \item Comment: ``ur my hero''; Label: \texttt{support}
    \item Comment: ``I’m so sorry!! Some people are just so disgusting. You’re awesome and your content is great. Please hang in there as we all think you’re wonderful!''; Label: \texttt{support}
    \item Comment: ``Try each look with reading glasses and see what happens''; Label: \texttt{neutral}
    \item Comment: ``are people in Japan like that?''; Label: \texttt{neutral}
    \item Comment: ``You're overreacting, it's not that serious''; Label: \texttt{oppose}
    \item Comment: ``if u not saying no name piss of''; Label: \texttt{oppose}
\end{itemize}
Important:
\begin{itemize}
    \item The stance is relative to the associated post author's expressed perspective or framing.
    \item Do not determine whether the post author's perspective is factually true.
    \item If the comment is sarcastic, classify by intended meaning.
    \item If the stance is unclear, choose \texttt{neutral}.
\end{itemize}
\end{itemize}

\begin{table}[ht]  
\centering
\caption{Mean scores for each LIWC emotion and Linguistic (sub)category per harassment type}
\label{tab:mean scores LIWC}
\resizebox{\columnwidth}{!}{%
\scriptsize
\setlength{\tabcolsep}{2.5pt}
\renewcommand{\arraystretch}{1.05}
\begin{tabular}{l *{10}{r}}
\toprule
\textbf{Harassment Type} & \textbf{Overall} &
\textbf{Sexual} & \textbf{Bullying} & \textbf{Verbal} & \textbf{Catcall.} & \textbf{Stalking} & \textbf{Law Enf.} & \textbf{Physical} & \textbf{Other} & \textbf{Gaslight} \\
\midrule
\multicolumn{11}{l}{\textbf{LIWC Emotions}} \\
\midrule
Positive Emotions & 1.87 & 2.14 & 1.44 & 2.03 & 3.05 & 0.77 & 0.98 & 1.23 & 2.02 & 1.56 \\
Negative Emotions & 1.46 & 1.69 & 1.11 & 1.38 & 1.16 & 1.27 & 1.13 & 1.06 & 1.33 & 1.03 \\
Anxiety & 0.38 & 0.43 & 0.17 & 0.31 & 0.37 & 0.57 & 0.48 & 0.15 & 0.49 & 0.31 \\
Anger & 0.28 & 0.26 & 0.26 & 0.37 & 0.25 & 0.17 & 0.16 & 0.22 & 0.39 & 0.25 \\
Sadness & 0.30 & 0.44 & 0.21 & 0.20 & 0.14 & 0.13 & 0.15 & 0.30 & 0.17 & 0.09\\
\midrule
\multicolumn{11}{l}{\textbf{LIWC Linguistic}} \\
\midrule
Function Words & 49.32 & 49.66 & 48.03 & 50.93 & 52.77 & 50.81 & 47.48 & 42.05 & 47.36 & 49.39 \\
Pronouns & 18.31 & 18.37 & 17.95 & 19.16 & 19.77 & 18.26 & 16.96 & 16.14 & 17.77 & 19.20 \\
\textit{Personal Pronouns} & 12.50 & 12.39 & 12.29 & 13.26 & 13.56 & 12.41 & 11.71 & 10.59 & 12.50 & 13.12 \\
$\bullet$ 1st Person Singular & 4.77 & 5.38 & 4.26 & 4.62 & 7.15 & 3.87 & 2.34 & 4.89 & 5.43 & 3.24 \\
$\bullet$ 1st Person Plural & 0.51 & 0.59 & 0.40 & 0.58 & 0.48 & 0.44 & 0.51 & 0.31 & 0.45 & 0.49 \\
$\bullet$ 2nd Person & 3.54 & 3.33 & 3.74 & 3.87 & 3.40 & 3.46 & 4.55 & 2.17 & 3.80 & 4.68 \\
$\bullet$ 3rd Person Singular & 2.20 & 1.83 & 2.35 & 2.61 & 1.24 & 3.10 & 1.71 & 1.71 & 1.18 & 2.48 \\
$\bullet$ 3rd Person Plural & 1.27 & 1.09 & 1.21 & 1.27 & 1.13 & 1.37 & 2.44 & 1.30 & 1.47 & 2.08 \\
\textit{Impersonal Pronouns} & 5.81 & 5.98 & 5.66 & 5.89 & 6.21 & 5.85 & 5.24 & 5.56 & 5.28 & 6.08 \\
Determiners & 12.21 & 12.18 & 11.85 & 12.55 & 13.61 & 12.82 & 11.79 & 10.51 & 11.50 & 12.79 \\
\textit{Articles} & 4.71 & 4.48 & 4.58 & 4.94 & 4.90 & 5.44 & 5.00 & 4.21 & 4.66 & 4.89 \\
\textit{Numbers} & 1.54 & 1.33 & 2.43 & 1.11 & 1.45 & 1.26 & 1.15 & 4.28 & 2.30 & 1.12 \\
Prepositions & 8.44 & 8.28 & 8.45 & 8.38 & 9.10 & 8.70 & 8.74 & 6.92 & 8.58 & 8.18 \\
Auxiliary Verbs & 9.73 & 9.80 & 9.65 & 9.96 & 9.66 & 10.30 & 9.49 & 8.40 & 9.67 & 10.08 \\
Adverbs & 5.76 & 6.20 & 5.24 & 6.10 & 6.16 & 6.07 & 4.64 & 4.73 & 5.16 & 4.60 \\
Conjunctions & 4.36 & 4.83 & 3.84 & 4.45 & 5.15 & 4.33 & 3.84 & 3.62 & 3.99 & 3.72 \\
Negations & 2.19 & 2.19 & 2.23 & 2.33 & 2.05 & 2.43 & 2.17 & 1.96 & 2.01 & 2.24 \\
Common Verbs & 18.11 & 17.90 & 17.83 & 18.95 & 18.17 & 19.68 & 18.24 & 15.82 & 16.99 & 18.42 \\
Common Adjectives & 6.58 & 7.30 & 5.50 & 6.77 & 6.77 & 6.45 & 6.23 & 4.74 & 6.58 & 5.95 \\
Quantities & 2.48 & 2.32 & 2.96 & 2.53 & 2.69 & 2.13 & 2.06 & 3.83 & 2.31 & 2.63 \\
\bottomrule
\end{tabular}
}%
\end{table}

\begin{table}[ht]  
\centering
\scriptsize
\setlength{\tabcolsep}{3pt}
\renewcommand{\arraystretch}{1.05}
\begin{tabular}{l *{10}{r}}
\toprule
\textbf{Harassment Type} & \textbf{Overall Average} &
\textbf{Sexual} & \textbf{Bullying} & \textbf{Verbal} & \textbf{Catcalling} & \textbf{Stalking} & \textbf{Law Enf.} & \textbf{Physical} & \textbf{Other} & \textbf{Gaslighting} \\
\midrule
\multicolumn{11}{l}{\textbf{Harassment Target}} \\
\midrule
\textit{Self} & -1.5 & 1.1 & 1.8 & -1.9 & 0.48 & 0.64 & 1.9 & 1.3 & -1.2 & 1\\
\textit{Other} & 1.9 & -1.6 & -1.7 & 2.3 & 0.47 & -0.16 & -2.1 & 0.17 & 1.4 & -0.66\\
\textit{Celebrity} & 0.34 & 1.5 & 0.52 & -0.05 & -0.71 & 0.62 & -0.65 & 2.3 & -0.62 & -0.21 \\
\bottomrule
\end{tabular}
\caption{Standard residuals for harassment targets by harassment type}
\label{tab:res_targets}
\end{table}

\noindent \textbf{Note:} Since Cramer's V is computed in relation to the relevant dataset's size, it is possible for it to be undefined even when there is a statistically significant result. Where this has occurred, we have denoted it as \textit{"not defined"} in Tables~\ref{tab:stats_loc_vs_type}--\ref{tab:emotions_vs_type}.

\begin{table}[ht]  
\centering
\scriptsize
\setlength{\tabcolsep}{4pt}
\renewcommand{\arraystretch}{1.1}
\begin{tabular}{l l r r}
\toprule
\textbf{Harassment Type 1} & \textbf{Harassment Type 2} & \textbf{$\chi^2$ (p-value)} & \textbf{Cramer's V (where available)} \\
\midrule
Other & Catcalling & 0.013314 & >0.6 \\
& Sexual & 0.000006 & >0.4 \\
& Law Enforcement & 0.022751 & not defined \\
& Verbal & 0.005119 & not defined \\
& Bullying & 0.000291 & >0.6 \\
& Stalking & 0.005133 & >0.6 \\
\bottomrule
\end{tabular}
\caption{$\chi^2$ test of independence between harassment type and location ($\alpha$ = 0.05). We report only the significant results here.}
\label{tab:stats_loc_vs_type}
\end{table}

\FloatBarrier

\begin{table}[ht]  
\centering
\scriptsize
\setlength{\tabcolsep}{3pt}
\renewcommand{\arraystretch}{1.05}
\begin{tabular}{l l r r}
\toprule
\textbf{Harassment Type 1} & \textbf{Harassment Type 2} & \textbf{$\chi^2$ (p-value)} & \textbf{Cramer's V (where available)} \\
\midrule
Sexual & Law Enforcement & 0.000097 & not defined \\
& Gaslighting & 0.001464 & >0.4 \\
\bottomrule
\end{tabular}
\caption{$\chi^2$ test of independence between harassment type and video type ($\alpha$ = 0.05). We report only the significant results here.}
\label{tab:stats_type_vs_type}
\end{table}

\begin{table}[ht]  
\centering
\scriptsize
\setlength{\tabcolsep}{3pt}
\renewcommand{\arraystretch}{1.05}
\begin{tabular}{l l r r}
\toprule
\textbf{Harassment Type 1} & \textbf{Harassment Type 2} & \textbf{$\chi^2$ (p-value)} & \textbf{Cramer's V (where available)} \\
\midrule
Sexual & Other & 0.000004 & >0.4 \\
& Verbal & 0.048002 & not defined \\
\midrule
Catcalling & Law Enforcement & 0.017007 & not defined \\
& Verbal & 0.000565 & not defined \\
& Bullying & 0004737 & >0.4 \\
\midrule
Other & Law Enforcement & 0.020372 & not defined \\
& Verbal & 0.005507 & not defined \\
& Bullying & 0.00094 & >0.6 \\
\bottomrule
\end{tabular}
\caption{$\chi^2$ test of independence between harassment type and video theme ($\alpha$ = 0.05). We report only the significant results here.}
\label{tab:stats_theme_vs_type}
\end{table}

\begin{table}[ht]  
\centering
\scriptsize
\setlength{\tabcolsep}{3pt}
\renewcommand{\arraystretch}{1.05}
\begin{tabular}{l l r r}
\toprule
\textbf{Harassment Type 1} & \textbf{Harassment Type 2} & \textbf{$\chi^2$ (p-value)} & \textbf{Cramer's V (where available)} \\
\midrule
Other & Sexual & 0.000134 & not defined \\
& Law Enforcement & 0.01972 & not defined \\
& Bullying & 0.0004817 & not defined \\
& Stalking & 0.001706 & not defined \\
\bottomrule
\end{tabular}
\caption{$\chi^2$ test of independence between harassment type and emotions conveyed ($\alpha$ = 0.05). We report only the significant results here.}
\label{tab:emotions_vs_type}
\end{table}

\begin{table}[ht]  
\centering
\scriptsize
\setlength{\tabcolsep}{3pt}
\renewcommand{\arraystretch}{1.05}
\begin{tabular}{l r r r}
\toprule
& \textbf{Support} & \textbf{Neutral} & \textbf{Oppose} \\
\midrule
\textbf{Percentage} & 35.7374\% & 47.8628\% & 16.3997\% \\
\midrule
\textbf{Count} & 23,720 & 31,768 & 10,885 \\
\bottomrule
\end{tabular}
\caption{Distribution of the stances across the 66,373 comments from the 190 videos. Stances were detected via a human--aligned GPT--4.1--mini model in January 2026.}
\label{tab:stance_distribution}
\end{table}

\FloatBarrier

\begin{table}[ht]  
\centering
\scriptsize
\setlength{\tabcolsep}{3pt}
\renewcommand{\arraystretch}{1.05}
\begin{tabular}{l r r r r r r r}
\toprule
\textbf{Harassment Type} & \textbf{Sexual} & \textbf{Bullying} & \textbf{Verbal}  & \textbf{Stalking} & \textbf{Law Enf.} & \textbf{Physical} & \textbf{Other} \\
\midrule
\multicolumn{8}{l}{\textbf{Stance}} \\
\midrule
\textit{Neutral} & 15,552 & 8,282 & 9,364 & 5,557 & 3,103 & 2,897 & 1,847 \\
\textit{Support} & 14,078 & 4,298 & 6,923 & 5,813 & 1,791 & 1,097 & 1,110 \\
\textit{Oppose} & 3,971 & 3,458 & 4,610 & 2,494 & 1,150 & 833 & 808 \\
\bottomrule
\end{tabular}
\caption{Stance distributions for each harassment type. Note that subcategories of sexual and verbal harassment were merged to avoid generating undue statistical noise or spuriously significant results.}
\label{tab:stance_by_type}
\end{table}

\begin{table}[ht]  
\centering
\scriptsize
\setlength{\tabcolsep}{3pt}
\renewcommand{\arraystretch}{1.05}
\begin{tabular}{l r r r r r r r}
\toprule
\textbf{Harassment Type} & \textbf{Sexual} & \textbf{Bullying} & \textbf{Verbal} & \textbf{Stalking} & \textbf{Law Enf.} & \textbf{Physical} & \textbf{Other} \\
\midrule
\multicolumn{8}{l}{\textbf{Stance}} \\
\midrule
\textit{Neutral}  & -2.06 & 8.46 & -4.73 & -11.97 & 4.86 & 13.13 & 1.79 \\
\textit{Support}  & 19.84 & -18.40 & -5.64 & 12.81 & -7.60 & -14.85 & -6.15 \\
\textit{Oppose}   & -24.87 & 12.32 & 15.79 & 1.40 & 2.85 & -0.39 & 5.82 \\
\bottomrule
\end{tabular}
\caption{Standardized residuals from the $\chi^2$ test of independence for stance by harassment type. Positive residuals indicate combinations occurring more frequently than expected under independence, while negative values indicate fewer occurrences than expected. Large absolute residuals highlight the primary contributors to the overall association between stance and harassment type.}
\label{tab:stance_residuals}
\end{table}

\begin{table}[ht]
\centering
\scriptsize
\setlength{\tabcolsep}{3pt}
\renewcommand{\arraystretch}{1.05}
\begin{tabular}{l r r r r}
\toprule
\textbf{Parameter} & \textbf{Mean} & \textbf{SD} & \textbf{HDI 3\%} & \textbf{HDI 97\%} \\
\midrule
\multicolumn{5}{l}{\textit{Intercepts}} \\
$\alpha_{\text{oppose}}$  & -1.39 & 0.38 & -2.13 & -0.73 \\
$\alpha_{\text{support}}$ & -0.72 & 0.38 & -1.46 & -0.03 \\
\midrule
\multicolumn{5}{l}{\textit{Harassment Effects (Oppose)}} \\
$\beta_{\text{oppose, Bullying}}$ & -0.10 & 0.43 & -0.87 & 0.75 \\
$\beta_{\text{oppose, Law Enforcement}}$ & 0.30 & 0.46 & -0.63 & 1.10 \\
$\beta_{\text{oppose, Other}}$ & -0.36 & 0.45 & -1.14 & 0.53 \\
$\beta_{\text{oppose, Physical}}$ & -0.71 & 0.59 & -1.77 & 0.37 \\
$\beta_{\text{oppose, Sexual}}$ & -0.25 & 0.40 & -0.95 & 0.51 \\
$\beta_{\text{oppose, Stalking/Staring}}$ & 0.26 & 0.47 & -0.60 & 1.18 \\
$\beta_{\text{oppose, Verbal}}$ & 0.31 & 0.43 & -0.53 & 1.04 \\
\midrule
\multicolumn{5}{l}{\textit{Harassment Effects (Support)}} \\
$\beta_{\text{support, Bullying}}$ & 0.05 & 0.42 & -0.74 & 0.83 \\
$\beta_{\text{support, Law Enforcement}}$ & -0.35 & 0.48 & -1.25 & 0.53 \\
$\beta_{\text{support, Other}}$ & -0.71 & 0.50 & -1.68 & 0.18 \\
$\beta_{\text{support, Physical}}$ & -0.97 & 0.56 & -2.05 & 0.06 \\
$\beta_{\text{support, Sexual}}$ & 0.40 & 0.38 & -0.33 & 1.10 \\
$\beta_{\text{support, Stalking/Staring}}$ & 0.71 & 0.45 & -0.02 & 1.72 \\
$\beta_{\text{support, Verbal}}$ & 0.32 & 0.44 & -0.54 & 1.14 \\
\midrule
\multicolumn{5}{l}{\textit{Random Effect Variance}} \\
$\sigma_{\text{video, oppose}}$ & 1.02 & 0.06 & 0.90 & 1.13 \\
$\sigma_{\text{video, support}}$ & 1.04 & 0.06 & 0.92 & 1.15 \\
\bottomrule
\end{tabular}
\caption{Posterior parameter estimates from the Bayesian hierarchical multinomial logistic regression. Coefficients are on the log-odds scale relative to the neutral reference category. Random-effect standard deviations quantify video-level variation in stance.}
\label{tab:hierarchical_params}
\end{table}

\begin{table}[ht]
\centering
\scriptsize
\setlength{\tabcolsep}{3pt}
\renewcommand{\arraystretch}{1.05}
\begin{tabular}{l c c c}
\toprule
\textbf{Harassment Type} & \textbf{Neutral} & \textbf{Oppose} & \textbf{Support} \\
\midrule
Bullying & 0.57 [0.50, 0.65] & 0.13 [0.09, 0.20] & 0.30 [0.22, 0.38] \\
Law Enforcement & 0.59 [0.49, 0.68] & 0.20 [0.12, 0.32] & 0.21 [0.12, 0.32] \\
Other & 0.70 [0.58, 0.79] & 0.13 [0.07, 0.22] & 0.17 [0.09, 0.29] \\
Physical & 0.75 [0.60, 0.87] & 0.10 [0.03, 0.22] & 0.15 [0.06, 0.31] \\
Sexual & 0.52 [0.47, 0.57] & 0.10 [0.08, 0.13] & 0.38 [0.32, 0.44] \\
Stalking/Staring & 0.43 [0.32, 0.53] & 0.14 [0.07, 0.24] & 0.43 [0.29, 0.57] \\
Verbal & 0.49 [0.40, 0.58] & 0.17 [0.11, 0.25] & 0.33 [0.24, 0.45] \\
\bottomrule
\end{tabular}
\caption{Posterior stance probabilities by harassment type estimated from the hierarchical multinomial model. Values represent posterior means, with 95\% highest density intervals shown in brackets. Probabilities are marginalized over video-level random effects.}
\label{tab:posterior_stance_probs}
\end{table}

\begin{table}[ht]
\centering
\scriptsize
\setlength{\tabcolsep}{3pt}
\renewcommand{\arraystretch}{1.05}
\begin{tabular}{l c c c}
\toprule
\textbf{Comparison} & \textbf{Mean Difference} & \textbf{95\% HDI} & \textbf{$P(\text{diff} > 0)$} \\
\midrule
Bullying $-$ Law Enforcement & 0.17 & [0.05, 0.30] & 1.00 \\
Bullying $-$ Other & 0.16 & [0.03, 0.28] & 1.00 \\
\bottomrule
\end{tabular}
\caption{Posterior contrasts in polarization (defined as $P(\text{support}) + P(\text{oppose})$) between harassment types. Only contrasts with credible non-zero differences are shown. Positive values indicate higher polarization in the first-listed category.}
\label{tab:polarization_contrasts}
\end{table}

\FloatBarrier

\begin{table}[H]
\centering
\scriptsize
\setlength{\tabcolsep}{3pt}
\renewcommand{\arraystretch}{1.05}
\begin{tabular}{l c c c}
\toprule
\textbf{Parameter} & \textbf{Mean} & \textbf{SD} & \textbf{95\% HDI} \\
\midrule
$\sigma_{\text{video, oppose}}$ & 1.02 & 0.06 & [0.90, 1.13] \\
$\sigma_{\text{video, support}}$ & 1.04 & 0.06 & [0.92, 1.15] \\
\bottomrule
\end{tabular}
\caption{Posterior standard deviations of video-level random effects from the hierarchical multinomial model. Values are on the log-odds scale and quantify the degree of contextual variation in stance across videos. Larger values indicate greater between-video heterogeneity in stance expression.}
\label{tab:video_random_effects}
\end{table}

\begin{table}[H]
\centering
\scriptsize
\setlength{\tabcolsep}{3pt}
\renewcommand{\arraystretch}{1.05}
\begin{tabular}{l r r r r r}
\toprule
\textbf{Predictor} & \textbf{Coef.} & \textbf{SE} & \textbf{z} & \textbf{p} & \textbf{95\% CI} \\
\midrule
Intercept  & 0.88 & 0.08 & 11.16 & $<$0.001 & [0.72, 1.03] \\
Emotion    & 0.24 & 0.06 & 4.32  & $<$0.001 & [0.13, 0.35] \\
Power/Agency & 0.16 & 0.03 & 5.24 & $<$0.001 & [0.10, 0.22] \\
Cognitive/Analytic & -0.16 & 0.03 & -4.70 & $<$0.001 & [-0.23, -0.09] \\
Social & 0.01 & 0.02 & 0.34 & 0.732 & [-0.04, 0.05] \\
Moral  & -0.01 & 0.02 & -0.49 & 0.625 & [-0.04, 0.02] \\
\bottomrule
\end{tabular}
\caption{Generalized estimating equation (GEE) logistic regression predicting supportive versus oppositional stance among engaged comments. Coefficients are on the log-odds scale. The model accounts for clustering of comments within videos using an exchangeable correlation structure and robust standard errors.}
\label{tab:gee_liwc}
\end{table}

\begin{table}[H]
\centering
\scriptsize
\setlength{\tabcolsep}{3pt}
\renewcommand{\arraystretch}{1.05}
\begin{tabular}{l r r r r}
\toprule
\textbf{LIWC Category} & \textbf{Mean (Stalking)} & \textbf{Mean (Physical)} & \textbf{P-Value} & \textbf{Cohen's d} \\
\midrule
adj        & 6.547 & 4.258 & 0.005216 & 1.286 \\
you        & 3.598 & 1.787 & 0.012375 & 1.112 \\
verb       & 19.560 & 14.407 & 0.011156 & 1.085 \\
function   & 50.694 & 38.324 & 0.035764 & 1.027 \\
prep       & 8.834 & 6.107 & 0.060038 & 1.026 \\
adverb     & 5.812 & 4.172 & 0.018512 & 1.008 \\
auxverb    & 10.261 & 7.603 & 0.029808 & 1.000 \\
det        & 12.649 & 9.471 & 0.120691 & 0.945 \\
ppron      & 12.536 & 9.903 & 0.039104 & 0.843 \\
conj       & 4.362 & 3.144 & 0.070687 & 0.781 \\
article    & 5.171 & 3.881 & 0.112197 & 0.781 \\
we         & 0.423 & 0.245 & 0.044460 & 0.778 \\
pronoun    & 18.500 & 15.070 & 0.129680 & 0.763 \\
negate     & 2.267 & 1.827 & 0.170904 & 0.464 \\
number     & 1.564 & 3.702 & 0.351787 & -0.438 \\
ipron      & 5.964 & 5.178 & 0.390838 & 0.405 \\
shehe      & 2.697 & 2.148 & 0.701419 & 0.254 \\
they       & 1.411 & 1.172 & 0.941778 & 0.182 \\
i          & 4.237 & 4.343 & 0.840830 & -0.073 \\
\bottomrule
\end{tabular}
\caption{Comparison of LIWC Linguistic subcategories between stalking and physical harassment videos. Categories are ordered by effect size (Cohen's d). Positive values indicate higher average scores for stalking videos, while negative values indicate higher averages for physical harassment videos.}
\label{tab:liwc_linguistic_stalking_physical}
\end{table}